\documentclass[11pt]{article} 

\usepackage[utf8]{inputenc} 

\usepackage{geometry} 
\usepackage{graphicx} 
\usepackage{hyperref}
\usepackage{graphicx}
\usepackage{rotating}
\usepackage{pdflscape}
\usepackage{amsmath}
\usepackage{amssymb}
\usepackage{amsthm}
\usepackage{epstopdf}
\usepackage{multirow}
\usepackage{stmaryrd}
\usepackage{bbm}
\usepackage{bm}
\usepackage[round]{natbib}

\newcommand{\indep}{\;\, \rule[0em]{.03em}{.6em}\hspace{-.25em}\rule[0em]{.65em}{.03em}\hspace{-.25em}\rule[0em]{.03em}{.6em}\;\,}

\newtheorem{definition}{Definition} 

\title{Bayesian Tensor Regression for Neuroimaging Data}
\author{Zahra Naji$^{1}$, 
Montserrat Fuentes$^{2}$, 
Liangsuo Ma$^{3}$, 
Hossein Moradi Rekabdarkolaee$^{4,*}$ \\
$^{1}$ Sanford Health, Sioux Fall, SD,\\
$^{2}$ St. Edward's University, Austin, TX, \\
$^{3}$ Virginia Commonwealth University, Richmond, VA, \\
$^{4}$ Department of Business Analytics, Economics, and Information Systems\\
Bowling Green State University, Bowling Green, OH.
}

\date{}
\begin{document}
\maketitle

\hrule

\begin{abstract}
Multidimensional array data, also called tensors, are used in neuroimaging and other big data applications. In this paper, we propose a parsimonious Bayesian Tensor linear model for neuroimaging study with brain image as a response and a vector of predictors. Our method provides estimates for the parameters of interest by using an Envelope method. The proposed method characterizes different sources of uncertainty and the inference is performed using MCMC. We demonstrate posterior consistency and develop a computationally efficient Markov Chain Monte Carlo algorithm for posterior computation using Gibbs sampling. The effectiveness of our approach is illustrated through simulation studies and analysis of Alcohol use disorder's effect on brain connectivity. 
\end{abstract}

\vspace{0.1in}

\hrule


\section{Introduction}
\label{intro}

Excessive alcohol use has numerous detrimental effects on physical and mental health and societal well-being \citep{rehm2009global}. According to the Centers for Disease Control and Prevention (CDC), alcohol-related deaths in the United States averaged over $95,000$ per year between 2011 and 2015, with approximately 261 deaths occurring daily. Additionally, excessive alcohol use is associated with numerous chronic health conditions, such as liver disease, cardiovascular problems, and certain types of cancer. It also increases the risk of mental health disorders, including depression and anxiety. Furthermore, alcohol abuse leads to substantial economic costs, estimated at \$249 billion in 2010, encompassing healthcare expenditures, criminal justice expenses, and reduced workplace productivity \citep{sacks2015}. Recent studies have shown that chronic alcohol consumption induces neuroinflammation and neurodegeneration, leading to significant brain damage \citep{crews2014neuroimmune, zahr2014structural}. Moreover, alcohol abuse has been associated with structural brain changes, including reduced gray matter volume and alterations in white matter integrity, which are linked to cognitive impairments and increased vulnerability to alcohol-related brain disorders \citep{ros2018alcohol}.

Imaging techniques are fundamental tools in advancing our understanding of brain functionality \citep{lindquist2008statistical, lazar2008statistical, zhou2013tensor, bellec2017neuro}. Electroencephalography (EEG), magnetic resonance imaging (MRI), functional magnetic resonance imaging (fMRI), and diffusion tensor imaging (DTI) are examples of such data. Imaging data are best represented in the form of a multidimensional array, also called a tensor \citep{kolda2009tensor}. These data provide access to brain activity at the voxel level. The complex structure of this type of data calls for flexible and sophisticated models to better understand brain activity. This paper proposes a Bayesian tensor response regression model to study brain images and their association with alcohol use.

The majority of the literature on brain image analysis falls into either ultrahigh-dimensional regression approaches \citep{helland1990partial, chun2010sparse, yuan2007dimension, chen2012sparse} or functional data analysis (FDA) approaches \citep{ramsay2006functional, reiss2010functional, jiang2016functional}. Recently, there have been developments in tensor regression models, where a multidimensional array is considered either a predictor or a response \citep{zhou2013tensor, zhou2014regularized, goldsmith2014smooth, wang2014regularized, zhang2017tensor, li2017parsimonious, deng2021tensor}. Predictor tensor methods use brain images as tensor-valued covariates to predict a scalar or vector-valued clinical outcome \citep{guhaniyogi2017bayesian, spencer2022parsimonious}. Tensor response methods study the association between an image-valued response and a vector of covariates to understand brain activity patterns across different groups of subjects \citep{li2017parsimonious, giefer2025employing}. Our proposed method falls into the tensor response category, with the fMRI image serving as the response variable. In addition to jointly modeling all voxels in an image, our approach is Bayesian, allowing for the incorporation of multiple sources of uncertainty.

This paper proposes a Bayesian tensor regression model to explore the association between an fMRI image response and a vector of predictors. Our approach provides an inferential framework for investigating the impact of specific covariates on brain images and for assessing whether these effects vary across the brain. The novel aspects of our proposed model are as follows. First, we utilize and extend a Bayesian dimension reduction approach, the Bayesian envelope \citep{khare2017bayesian}, to a tensor response linear regression setting. This dimension reduction technique assumes the existence of linear combinations of response variables that are irrelevant to the regression, referred to as the immaterial part of the regression. By eliminating this immaterial component, the variance of parameter estimation can be reduced. This approach naturally leads to a tensor Tucker decomposition \citep{li2017parsimonious}. Second, our model jointly considers all voxels in the brain image, thereby accounting for spatial correlation in the data. Third, parameter estimation is performed using a Gibbs sampling algorithm, ensuring computational feasibility. The Bayesian framework further enables uncertainty quantification through posterior distributions. Finally, we study the posterior consistency of the proposed model.

The remainder of the paper is organized as follows. Section \ref{summary} provides a description of the data. Section \ref{brief} provides a brief review for tensor notation. The proposed Bayesian tensor regression model is detailed in section \ref{method}. Sections \ref{sim} and \ref{real} contain simulation studies and analyses of the association between brain activity and demographic information and Alcohol use disorder. We conclude the paper with a short discussion and suggestions for future work in Section \ref{conc}. The theoretical results and posterior specifications for different models used in the paper are presented in the Appendix.


\section{Data}
\label{summary}

The data analyzed in Section \ref{real} were obtained from the Human Connectome Project (HCP) Young Adult study. The HCP data are freely available at \url{www.humanconnectome.org/study/hcp-young-adult} and include group-average structural and functional MRI data. All experimental protocols for the HCP project were approved by the Institutional Review Board (IRB \#201204036; Title: ``Mapping the Human Connectome: Structure, Function, and Heritability''). We focus on the alcohol use subset of the data, which contains 178 subjects: 89 with alcohol use and 89 non--alcohol users. The sample includes 90 males and 88 females, with approximately 72\% of subjects identifying as White, not Hispanic/Latino. Each fMRI image is a three-dimensional array with dimensions $145 \times 174 \times 145$. 

The nature of the data presents several challenges. Jointly modeling all voxels in an fMRI image results in an extremely large number of parameters (over 7 million voxels), necessitating a sophisticated dimension reduction approach to reduce dimensionality while preserving essential information. Furthermore, because the data are highly noisy, dimension reduction improves the efficiency of parameter estimation. Finally, multiple sources of uncertainty must be accounted for in the modeling. To address these challenges, we propose a Bayesian tensor regression model with vector-valued covariates.


\section{Breif Review}
\label{brief}
This section briefly reviews the tensor notation used throughout the paper, a linear model with a tensor response and vector covariates, and the tensor envelope methodology. 

\subsection{Tensor Notation}
A multidimensional array $\mathbbm{A}\in \mathbbm{R}^{r_1\times\ldots\times r_m}$, called an $m$-th order tensor, is a  generalization of vectors and matrices. For a tensor $\mathbbm{A}$, $\mathbbm{A}_{(k)}$ denotes the mode-k matricization that maps the tensor into a matrix, i.e.  $\mathbbm{A}_{(k)} \in \mathbbm{R}^{r_k \times (\Pi_{j\neq k} r_{j})}$. The \textit{k-mode} product of a tensor $\mathbbm{A}$ and a matrix $\textbf{C} \in \mathbbm{R}^{s\times r_k}$, shown by $\mathbbm{A} \times_{k} \textbf{C} \in \mathbbm{R}^{r_1 \times \ldots r_{k-1} \times s \times r_{k+1}\ldots\times r_m}$, is when mode-k of $\mathbbm{A}$ is multiplied by the matrix $\textbf{C}$. 
For a tensor $\mathbbm{A}$, we let $\mathbbm{A} = \llbracket \mathbbm{a}; \textbf{G}_{1},\ldots, \textbf{G}_{m} \rrbracket$ denote the tucker decomposion of the tensor with $\mathbbm{a}$ being the core tensor, and, for $k=1,\ldots,m$, $\textbf{G}_{k}$ are the, often orthogonal, factor matrices. A detailed review of tensors and their properties can be found in \cite{kolda2009tensor} and the references therein. 


\subsection{Linear Model with Tensor Response}  
Let $\mathbbm{Y}_{i}\in \mathbbm{R}^{r_1\times \ldots \times r_m}$ be the order $m$ tensor, for $i=1,\ldots,n$, showing the response variable. The tensor response linear model is 
\begin{equation}
\label{model1}
\mathbbm{Y}_{i} =  \mathbb{\beta} \times_{(m+1)}\textbf{x}_{i}+\mathbb{\epsilon}_{i},
\end{equation}
where $\textbf{x}_{i}$ denotes the $1\times p$ vector of non-random predictors, $ \mathbb{\beta}\in \mathcal{R}^{r_1\times \ldots \times r_m\times p}$ is an $(m + 1)$-th order tensor of the coefficients, and $\mathbb{\epsilon}_i$ denotes the error which is an $m$-th order tensor that is independent of $\textbf{x}_i$ and has mean zero. We assumed  $\mathbb{\epsilon}_i$ has a separable covariance structure such that $cov(vec(\mathbb{\epsilon})) = \boldsymbol\Sigma = \boldsymbol\Sigma_{m} \otimes \ldots \otimes \boldsymbol\Sigma_{1}$ where $\boldsymbol\Sigma_{k}$s are positive definite for $k=1,\ldots,m$ and $\otimes$ denotes the Kronecker product. The goal is to estimate the unknown coefficients $\mathbb{\beta} $. Through out this paper, we assume $\mathbbm{Y}_{i}$ follows a tensor normal distribution of order m, which is defined as: 
\[
\mathbbm{Y}_{i} \sim TN( \mathbb{\beta} \times_{(m+1)}\textbf{x}_{i}; \boldsymbol\Sigma_{1}, \boldsymbol\Sigma_{2},\ldots,\boldsymbol\Sigma_{m}).
\]
It is worth mentioning that a random variable such as $\mathbbm{A}\in \mathcal{R}^{r_1\times\ldots\times r_m}$ follows a tensor normal distribution of order m with mean $\mathbb{\mu}$ and covariance $[\boldsymbol\Sigma_{1}, \boldsymbol\Sigma_{2},\ldots,\boldsymbol\Sigma_{m}]$ if and only if the vectorized version of the $\mathbbm{A}$ follows the following normal distribution 
\[
vec(\mathbbm{A}) \sim N_{\prod_{i=1}^{m} r_i} (vec(\mathbb{\mu}),\boldsymbol\Sigma_{m} \otimes \ldots \otimes \boldsymbol\Sigma_{1}).
\] 
\cite{manceur2013maximum} provided a computationally efficient algorithm for maximum likelihood estimation for the tensor linear model. 
There are two major low-rank decompositions for tensor regression to reduce the ultrahigh dimensional estimation problem to a manageable size problem. \cite{zhou2013tensor} proposed using CP (CANDECOMP/PARAFAC) decomposition of the regression coefficient array and \cite{li2018tucker} proposed using Tucker decomposition for this problem. \cite{li2018tucker} showed that while the Tucker decomposition approach is computationally more expensive compared to CP, it provides more accurate estimation results and has a smaller number of free parameters.

\subsection{Tensor Envelope}
We borrowed the notation from \cite{li2017parsimonious} to present the Tensor Envelope for the ease of reader. \textit{Envelope} is a parsimonious approach for multivariate response linear model based on the assumption that some linear combinations of response do not depend on any of the predictors \citep{cook2010envelope}. The goal of the envelope is to find an orthogonal matrix $(\boldsymbol\Gamma_{1}, \boldsymbol\Gamma_{0}) \in \mathcal{R}^{r\times r}$ that decompose $\textbf{Y}$ into $\textbf{P}_{\boldsymbol\Gamma_1}\textbf{Y}$ and $\textbf{Q}_{\boldsymbol\Gamma_1}\textbf{Y}$, where $\textbf{P}_{\boldsymbol\Gamma_1} = \boldsymbol\Gamma_{1}\boldsymbol\Gamma_{1}^{T}$ is an orthogonal projection operator with respect to the standard inner product, and $\textbf{Q}_{\boldsymbol\Gamma_1}=\textbf{I}_{r}-\textbf{P}=\boldsymbol\Gamma_{0}\boldsymbol\Gamma_{0}^{T}$ is the projection onto the complement space. Suppose this orthogonal matrix satisfies the following two conditions: (i) $span(\boldsymbol\beta) \subseteq span(\boldsymbol\Gamma_{1})$, and (ii) $\textbf{P}_{\boldsymbol\Gamma_1}\textbf{Y}$ is conditionally independent of $\textbf{Q}_{\boldsymbol\Gamma_1}\textbf{Y}$ given \textbf{X}. \cite{cook2010envelope} showed that using these two conditions, one can decompose the covariance matrix  of response into the covariance of the material part, $\boldsymbol\Sigma_{1}= var(\textbf{P}_{\boldsymbol\Gamma_1}\textbf{Y}|\textbf{X})$, and immaterial part, $\boldsymbol\Sigma_{2}= var(\textbf{Q}_{\boldsymbol\Gamma_1}\textbf{Y}|\textbf{X})$. Using this decomposition the covariance function of response becomes $\boldsymbol\Sigma= \boldsymbol\Sigma_{1} + \boldsymbol\Sigma_{2}$ and the goal is to find the subpace $\mathcal{E}=span(\boldsymbol\Gamma_1)$.  \cite{li2017parsimonious} extended this idea to higher order tensor. For a tensor of the order of $m$ assume there exist a series of subspaces, $\mathcal{E}_{k} \subseteq  \mathcal{R}^{r_{k}}$ for $k=1,\ldots,m$, such that
\begin{eqnarray}
\label{eq1}
&\mathbbm{Y}\times_{k}\textbf{Q}_{k}| \textbf{X} \sim\mathbbm{Y}\times_{k}\textbf{Q}_{k},\cr
&\mathbbm{Y}\times_{k}\textbf{Q}_{k}\indep \mathbbm{Y}\times_{k}\textbf{P}_{k} | \textbf{X},
\end{eqnarray}
where $\indep$ denotes the statistical independence, $\textbf{P}_{k}$ shows an orthogonal projection operator to subspace $\mathcal{E}_{k}$ with respect to the standard inner product, and $\textbf{Q}_{k}=\textbf{I}_{r_{k}}-\textbf{P}_{k}$ denotes the projection onto the complement space of $\mathcal{E}_{k}$. $\boldsymbol\Gamma_{1}\in \mathcal{R}^{r_{k}\times u_{k}}$ is a basis matrix of $\mathcal{E}_{k}$, and $u_{k}\leq r_{k}$ is the dimension of $\mathcal{E}_{k}$ referred as dimension of the \textit{envelope subspace}. Combining the statements in (\ref{eq1}) for all $k = 1, \ldots, m$, we have a parsimonious representation:
\begin{eqnarray}
\label{eq2}
&\mathbbm{Q}(\mathbbm{Y})| \textbf{X} \sim \mathbbm{Q}(\mathbbm{Y}),\cr
&\mathbbm{Q}(\mathbbm{Y}) \indep \mathbbm{P}(\mathbbm{Y}) | \textbf{X},
\end{eqnarray}
where $\mathbbm{Q}(\mathbbm{Y}) = \mathbbm{Y} - \mathbbm{P}(\mathbbm{Y})\in \mathcal{R}^{r_1 \times \ldots \times r_m}$ and $\mathbbm{P}(\mathbbm{Y}) = \llbracket \mathbbm{Y}; \textbf{P}_1,\ldots, \textbf{P}_m \rrbracket$ is a \textit{Tucker decomposition} with $\mathbbm{Y}$ as core tensor and $\textbf{P}_1,\ldots, \textbf{P}_m$ as the factor matrices along with each mode. 

Using this methodology for decomposition of the covariance matrix for $k=1,\ldots,m$, we can write the covariance matrix as
\begin{eqnarray}
\label{eq3}
\boldsymbol\Sigma_{k}&=&\textbf{P}_{k}\boldsymbol\Sigma_{k} \textbf{P}_{k}+\textbf{Q}_{k}\boldsymbol\Sigma_{k} \textbf{Q}_{k}\cr
&=& \boldsymbol\Gamma_{1k}\boldsymbol\Omega_{1k}\boldsymbol\Gamma_{1k}^{T} + \boldsymbol\Gamma_{0k}\boldsymbol\Omega_{0k}\boldsymbol\Gamma_{0k}^{T} 
\end{eqnarray}
where $\boldsymbol\Gamma_{1k}\in \mathbbm{R}^{r_{k}\times u_{k}}$ is the basis for the $\mathcal{E}_{k}$ with dimension $u_{k}$ and $\boldsymbol\Gamma_{0k}\in \mathbbm{R}^{r_{k}\times (r_{k}-u_{k})}$ represents the basis for the complement space of $\mathcal{E}_{k}$. Let $\boldsymbol\Omega_{1k} \in\mathbbm{R}^{u_{k}\times u_{k}}$ and $\boldsymbol\Omega_{0k} \in \mathbbm{R}^{(r_{k}-u_{k})\times (r_{k}-u_{k})}$ denotes two symmetric positive definite matrices and $\textbf{P}_{k} = \boldsymbol\Gamma_{k}\boldsymbol\Gamma_{k}^{T}$ and $\textbf{Q}_{k} = \boldsymbol\Gamma_{0k}\boldsymbol\Gamma_{0k}^{T}$. Furthermore, since $\textbf{Q}_{k}$ does not contain any regression information, $\mathbb\beta \times_{k} \textbf{Q}_{k} = 0$. Thus, we can rewrite the coefficients tensor as 
$\mathbb{\beta} = \llbracket \mathbb{\eta}; \boldsymbol\Gamma_{11},\ldots, \boldsymbol\Gamma_{1m},\textbf{I}_{p}\rrbracket$,
for some $\mathbb{\eta}\in \mathbbm{R}^{u_1\times\ldots\times u_m\times p}$. More details on the tensor envelope can be found in \cite{li2017parsimonious} and the references therein.


\section{Methodology}
\label{method}
This section details our methodology, prior specification, and posterior consistency for our proposed model. 
\subsection{Reparametrization}
The tensor envelope model proposed by \cite{li2017parsimonious} does not provide a unique specification of $\boldsymbol\Gamma_{1k}$, for $k=1,2,\ldots,m$. The basis $\boldsymbol\Gamma_{1k}$ can be any orthonormal basis of the envelope subspace i.e. not identified up to an orthogonal rotation. However, to have an effective Bayesian analysis, uniqueness is critical. In this section, we present a reparametrization for the tensor envelope method that provides a unique representation for $\boldsymbol\Gamma_{1k}$ by extending the reparametrization proposed by \cite{khare2017bayesian}, \cite{leebayesian}, and \cite{chakraborty2024comprehensive} to the tensor response. This unique representation allows the incorporation of prior information on the envelope subspace into the specification of hyperparameters.

In specific, for $k = 1,\ldots, m$, in equation (\ref{eq3}), $\boldsymbol\Omega_{1k}$ and $\boldsymbol\Omega_{0k}$ are full symmetric matrices. Therefore, using spectral decomposition, $\boldsymbol\Omega_{1k}$ and $\boldsymbol\Omega_{0k}$ can be rewritten as $\textbf{W}_{1k} \boldsymbol\Delta_{1k} \textbf{W}_{1k}^{T}$ and $\textbf{W}_{0k} \boldsymbol\Delta_{0k} \textbf{W}_{0k}^{T}$, respectively, where $\textbf{W}_{0k}$ and $\textbf{W}_{1k}$ are orthonormal matrices and $\boldsymbol\Delta_{1k}=diag(\omega_{1k,1}\ldots,\omega_{1k,u_k})$ and $\boldsymbol\Delta_{0k}=diag(\omega_{0k,1}\ldots,\omega_{0k,r_k-u_k})$ are diagonal matrix and their diagonal entries are arranged in decreasing order. Let  $\boldsymbol\Omega_{1k}^{(new)} = \boldsymbol\Delta_{1k}$, $\boldsymbol\Omega_{0k}^{(new)}=\boldsymbol\Delta_{0k}$, $\boldsymbol\Gamma_{1k}^{(new)}=\boldsymbol\Gamma_{1k}\textbf{W}_{1k}$, and $\boldsymbol\Gamma_{0k}^{(new)}=\boldsymbol\Gamma_{0k}\textbf{W}_{0k}$ (with changes of signs for columns whose maximum entry is not positive) as the ``new'' $\boldsymbol\Gamma_{1k}$ and $\boldsymbol\Gamma_{0k}$, respectively. Then, $\boldsymbol\Gamma_{1k}^{(new)}$ and $\boldsymbol\Gamma_{0k}^{(new)}$ belong to $S_{r_{k},u_{k}}^{+}$ and $S_{r_{k},r_{k} - u_{k}}^{+}$, respectively, where $S_{a,b}^{+}$ denotes the Stiefel Manifold which is the collection of all $a\times b$ semi-orthogonal matrices such that the maximum entry (in absolute value) for each column of the matrix is positive. Although this parametrization look similar to the original envelope, the $\boldsymbol\Gamma_{1k}^{(new)}$ is the unique orthogonal basis that belongs to Stiefel manifold and is orthogonal to $\boldsymbol\Gamma_{0k}^{(new)}$ i.e. $\boldsymbol\Gamma_{0k}^{(new)^{T}}\boldsymbol\Gamma_{1k}^{(new)}=\textbf{0}$. Based on this reparametrization, we can assign a prior distribution to the parameters of the model and drive posterior distribution for the parameters in our model. To simplify the notation, we omit the subscript ``new" for the rest of the paper. 


\subsection{Prior Distributions}
\label{prior}

Our general strategy for the prior specification is to find proper, yet relatively non-informative, distributions. That is, we seek to specify a prior distribution that spreads its density over a reasonable (practical) range of values while letting the data do most of the informing of the posterior distribution. Furthermore, we assumed 
$(\omega_{1k},\omega_{0k})$ and $\textbf{O}_{k}=[\boldsymbol\Gamma_{1k}~~\boldsymbol\Gamma_{0k}]$, for $k=1,\cdots, m$, are priori independent. The prior distributions of the model parameters are chosen as:
\[
\mathbb\eta|\boldsymbol\Gamma_{0k},\boldsymbol\Gamma_{1k},\boldsymbol\Omega_{0k},\boldsymbol\Omega_{1k} \sim TN(\mathbb\eta_0, \boldsymbol\Omega_{11},\ldots,\boldsymbol\Omega_{1m}, I_p),
\]
where $\mathbb\eta_0= \llbracket \mathbbm{B}; \boldsymbol\Gamma_{11}^T,\ldots, \boldsymbol\Gamma_{1m}^T,\textbf{I}_{p}\rrbracket$ with $\mathbbm{B}$ is the tensor mean for the prior of the regression coefficients, and $TN(\cdot)$ denotes the tensor normal distribution. 
For $k=1,\ldots,m$, $\textbf{O}_{k}=[\boldsymbol\Gamma_{1k}, \boldsymbol\Gamma_{0k}]$ follows a matrix Bingham distribution, denoted by
\[
\textbf{O}_{k}=[\boldsymbol\Gamma_{1k}, \boldsymbol\Gamma_{0k}] \propto \exp \left\{-\frac{1}{2}tr(\textbf{B}_k \textbf{O}_{k}^{T} \textbf{A}_k \textbf{O}_{k} )\right\},
\]
where for $k=1,\ldots,m$, $\textbf{A}_k$ and $\textbf{B}_k$ are positive definite matrices. 
The diagonal entries of $\boldsymbol\Omega_{0k}$ and $\boldsymbol\Omega_{1k}$ i.e. $(\omega_{0k}$ and $\omega_{1k})$ are assumed to follow the inverse-Gamma distribution with positive hyper parameters $\alpha_{0k}, \lambda_{0k}, \alpha_{1k}$, and $\lambda_{1k}$, respectively. In addition, since the entries of $\omega_{1k}$ are assumed to be order statistics of $u_{k}$ i.i.d. observations following Inverse-Gamma$(\alpha_{1k}, \lambda_{1k}, 0,\infty)$ distribution. The entries of $\omega_{0k}$ are priori distributed as order statistics of $r_{k} - u_{k}$ i.i.d. observations from the Inverse-Gamma$(\alpha_0k, \lambda_{0k}, 0,\infty)$ distribution. We let the entries of $\omega_{1k}$ and $\omega_{0k}$ to be independent from each other. Therefore, the joint prior density of $\omega_{1k}$ and $\omega_{0k}$ is proportional to
\[
\pi(\omega_{1k},\omega_{0k}) \propto \prod_{i=1}^{u_k} \omega_{1k,i}^{-\alpha_k-1}\exp\left\{ -\frac{\lambda_k}{\omega_{1k,i}}\right\} \prod_{i=1}^{r_k-u_k} \omega_{0k,i}^{-\alpha_{0k}-1}\exp\left\{ -\frac{\lambda_{0k}}{\omega_{0k,i}}\right\},
\]
where vectors $\omega_{1k}$ and $\omega_{0k}$ are in decreasing order. These conditionally conjugate priors for the parameters fastens the computation. The posterior distribution for the parameters is proportion to
\begin{eqnarray}
\label{eq8} 
\pi(\phi|y)\propto L(y|\phi)\pi(\phi),
\end{eqnarray}
where $\phi = (vec(\mathbb\eta), vec(\boldsymbol\Gamma_{0k}), vec(\boldsymbol\Gamma_{1k}), \omega_{0k}, \omega_{1k}; k=1,\ldots,m)$ refers to the all the parameters in the model, $L(\cdot)$ denotes the likelihood, and $\pi(\cdot)$ shows the prior distributions. We utilize a MCMC algorithm to obtain a sample from the joint posterior distribution. 
Then, we make inference about model parameters using a hybrid Gibbs sampling algorithm. Derivation of the posteriors for each parameter can be found in the Appendix, section \ref{tenp}.


\subsection{Posterior Consistency}
\label{asym}

This section establishes convergence results for our proposed tensor regression model. For simplicity, we assumed the intercept is omitted by centering the response and covariates. Suppose the data generating model is in the assumed model class (\ref{model1}) and for $k = 1,\ldots,m$, the envelope structural dimension $u_k$ is known. The Kulback-Leibler (KL) neighborhood around the true tensor $\mathbb\eta^{True}$ is defined as
\begin{equation*}
\underline{\mathbb\eta_{n}} = \left\{\mathbb\eta: \frac{1}{n} \sum_{i=1}^{n} KL(f(\mathbbm{Y}_{i}|\boldsymbol\eta^{True}),f(\mathbbm{Y}_{i}|\mathbb\eta))<\epsilon \right\}.
\end{equation*}
Let $\pi$ and $\Pi$ show the prior and the posterior distributions for the $n$ observations, respectively. Then 
\begin{equation*}
\Pi(\underline{\mathbb\eta_{n}} ) = \frac{\int_{\underline{\mathbb\eta_{n}} }f(\mathbbm{Y}|\mathbb\eta)\pi(\mathbb\eta)}{\int f(\mathbbm{Y}|\mathbb\eta)\pi(\mathbb\eta)},
\end{equation*}
where $f(\mathbbm{Y}|\mathbb\eta)$ denotes the density of $\mathbbm{Y}$ under the Tucker regression model. The posterior consistency is established by showing that
\begin{equation*}
\Pi(\underline{\mathbb\eta_{n}}) \rightarrow 1 ~\text{under}~\mathbb\eta^{True}~ \text{almost surely as} ~n\rightarrow \infty.
\end{equation*}

\textbf{Theorem 1:}  Let $M_n=\frac{tr(X^TX)}{n}$. For any $\epsilon>0$, the model is  posterior consistent under $\mathbb\eta^{True}$, when prior $\pi_{n}(\mathbb\eta)$ satisfies at
\begin{eqnarray}
\pi\left(\boldsymbol\eta: tr\left[(\mathbb\eta - \mathbb\eta^{True}) \boldsymbol\Omega_{1}^{-1} (\mathbb\eta - \mathbb\eta^{True})^{T} \right] < \frac{2\epsilon}{M_n} \right)>0,~~for~all ~large ~n.
\end{eqnarray}
Following \cite{pham2015trace}, for our prior, $T=tr\left[ (\mathbb\eta - \mathbb\eta^{True}) \boldsymbol\Omega_{1}^{-1} (\mathbb\eta_{n}-\mathbb\eta_{n}^{True})^{T}\right]$ follows a noncentral $\chi^{2}$ distribution with ${p\prod u_{i}}$ degrees of freedom and the non-centrality parameter is $\tau^{2} = tr\left[ (\mathbb\eta_0-\mathbb\eta^{True}) \boldsymbol\Omega_{1}^{-1} (\mathbb\eta_0-\mathbb\eta^{True})^{T}\right]$. Therefore, if $\frac{2\epsilon}{M_n}>0$ then $\pi\left(T < \frac{2\epsilon}{M_n} \right)>0$  which means the model is posterior consistent if, for large $n$, $M_n$ is positive and bounded. Proof of the theorem can be found in the Appendix, section \ref{proofcons}.

\section{Simulation}
\label{sim}
In this section, we carry out a simulation study to evaluate the finite-sample performance of the proposed Bayesian Tensor Regression model method for various dimensions. We studied the performance of our model under three different simulation scenarios: a case with no dimension reduction, a case where the dimension of the tensor is changing and the dimension of the envelope is fixed, and a case where the dimension of the tensor is fixed and the dimension of the envelope is changing. For all scenarios, the data $\{\mathbb{Y}_i, \textbf{x}_i\}_{i = 1}^n$, is generated from model (\ref{model1}). The predictor, $\textbf{x}_i$,  is a two-dimensional vector following a multivariate normal distribution with a mean of zero and covariance $\textbf{I}_{2\times 2}$. The tensor response, $\mathbb{Y}_i$, is a three-dimensional tensor.  The resulting tensor coefficient is a tensor of order-4 and the sample size is n = 30 and 50. The elements in $\boldsymbol\eta$ are sampled from a standard tensor-normal distribution with mean \textbf{0} and covariance $\textbf{I}_2\otimes\textbf{I}_{u_3}\otimes\textbf{I}_{u_2}\otimes\textbf{I}_{u_1}$. We let $\epsilon$ to follow $N(0,\boldsymbol\Sigma_{\epsilon})$, where $\boldsymbol\Sigma_{\epsilon} = \boldsymbol\Sigma_{3}\otimes \boldsymbol\Sigma_{2}\otimes \boldsymbol\Sigma_{1}$. For $k=1, 2, 3$, we let $\boldsymbol\Sigma_{k}=\left(\boldsymbol\Gamma_{1k}\boldsymbol\Omega_{1k}\boldsymbol\Gamma_{1k}^{T} + \boldsymbol\Gamma_{0k}\boldsymbol\Omega_{0k}\boldsymbol\Gamma_{0k}^{T}\right)$ where the matrix $(\boldsymbol\Gamma_{1k};\boldsymbol\Gamma_{0k})$ is obtained by orthogonalizing a $r_k \times r_k$, matrix of random uniform $(0, 1)$ variables. The matrices $\boldsymbol\Omega_{1k}$ and $\boldsymbol\Omega_{0k}$ are specified in each scenario. 

We compared the proposed model with a Bayesian tensor linear model with two different priors; the Bayesian OLS estimator and the Bayesian GLS estimator. It is worth mentioning, for both Bayesian OLS and GLS, observations are independent of each other.  For the Bayesian OLS estimator, we let the coefficients follow a tensor normal prior i.e. $\pi(\mathbb\beta)\sim TN(0,\sigma_1^2\textbf{I}_{1},\sigma_2^2\textbf{I}_{2},\sigma_3^2\textbf{I}_{3},\textbf{I}_p)$. This method does not consider any dependency for the regression coefficients. For the Bayesian GLS estimator, we let the coefficients to follow a tensor normal prior i.e. $\pi(\mathbb\beta)\sim TN(0,\boldsymbol\Sigma_{1},\boldsymbol\Sigma_{2}, \boldsymbol\Sigma_{3},\textbf{I}_p)$. This method considers the dependency between the regression coefficients. For $k=1,2,3$, we assumed $\boldsymbol\Sigma_{k}$ follows the inverse-Wishart distribution. The derivation of the posterior distribution for this model is provided in the Appendix sections \ref{olsp} and \ref{glsp} for Bayesian OLS and Bayesian GLS approaches, respectively. With these conjugate prior distributions, Markov chain Monte Carlo (MCMC) sampling using the Gibbs algorithm is employed to make the inference. We draw 1500 samples and discard the first 1000 as burning. Convergence is monitored using trace plots of several representative parameters. For each combination, we simulate 100 data replications and report the average and the standard error of computation time in seconds, $||\mathbb\beta - \hat{\mathbb\beta}||$, and the average interval width of the 95\% credible interval. All the computations for this simulation study were done using \textit{R-4.1.2 x64}, on a \textit{Linux Machine with IntelCore i7 CPU 2.79GHz and 16Gb RAM}. We reported the average running time in seconds.

\textit{First Scenario -- no dimension reduction:} In this scenario, we considered a model where the response is an order-3 tensor with response dimensions $(r_1, r_2, r_3) = (20,20,20)$ and $(50, 50, 50)$. For this case, for $k=1,2,3$, we have $u_k=r_k$ and $\boldsymbol\Sigma_{k}=\boldsymbol\Gamma_{1k}\boldsymbol\Omega_{1k}\boldsymbol\Gamma_{1k}^{T}$. For this scenario, we let $\boldsymbol\Omega_{1k}$ to be a $r_k\times r_k$ identity matrix.

\begin{small}
\begin{table}[htp]
\caption{ Average and Standard Deviation of $||\mathbb\beta-\hat{\mathbb\beta}||$ and Average Interval Width (AIW) for 95\% credible interval for an order-3 tensor with response dimensions $(r_1, r_2, r_3) = (20,20,20)$ and $(50, 50, 50)$}
\begin{center}
\label{tab1}
\begin{tabular}{|l|l|l| l| l| l| l| l| } \hline
 \multicolumn{3}{|c|}{ }    &  \multicolumn{2}{c|}{r=(20,20,20) }       &  \multicolumn{2}{c|}{r=(50,50,50)} \\
\hline 
 \multicolumn{2}{|c|}{ } 	&	n	&	30	&	50	&	30	&	50	\\		
\hline															
BTR-OLS	&	$||\mathbb\beta-\hat{\mathbb\beta}||$ 	&	mean	&	0.014	&	0.021	&	0.021	&	0.020	\\
	&		&	sd	&	0.002	&	0.002	&	0.002	&	0.002	\\
	&	AIW	&	mean	&	4.971	&	7.770	&	7.770	&	7.774	\\
	&		&	sd	&	0.079	&	0.050	&	0.050	&	0.077	\\
	&	time	(in sec.)&	mean	&	64.508	&	1565.429	&	1565.429	&	2524.222	\\
	&		&	sd	&	0.926	&	14.791	&	14.791	&	390.285	\\
\hline													
BTR-GLS	&	$||\mathbb\beta-\hat{\mathbb\beta}||$ 	&	mean	&	0.011	&	0.010	&	0.010	&	0.009	\\
	&		&	sd	&	0.003	&	0.002	&	0.002	&	0.002	\\
	&	AIW	&	mean	&	1.139	&	1.125	&	1.125	&	1.127	\\
	&		&	sd	&	0.041	&	0.038	&	0.038	&	0.032	\\
	&	time	(in sec.)&	mean	&	180.132	&	4028.987	&	4028.987	&	6710.228	\\
	&		&	sd	&	0.900	&	57.520	&	57.520	&	682.470	\\
\hline													
BTR-ENV	&	$||\mathbb\beta-\hat{\mathbb\beta}||$ 	&	mean	&	0.010	&	0.010	&	0.010	&	0.009	\\
	&		&	sd	&	0.002	&	0.002	&	0.002	&	0.002	\\
	&	AIW	&	mean	&	1.554	&	1.535	&	1.535	&	1.553	\\
	&		&	sd	&	0.120	&	0.075	&	0.075	&	0.123	\\
	&	Time (in sec.)	&	mean	&	552.252	&	12856.977	&	12856.977	&	19279.797	\\
	&		&	sd	&	3.226	&	36.641	&	36.641	&	2210.935	\\
\hline
\end{tabular}
\end{center}			
\end{table}
\end{small}

Table \ref{tab1} shows the results of the simulation for scenarios where the structural dimension of the envelope and the dimension of the tensor response are equal i.e. no dimension reduction. The Bayesian OLS approach outperforms other models in terms of computation for different scenarios, and our proposed method is the slowest. Furthermore, it can be observed that our model provides comparable results in terms of the average and the standard error of $||\mathbb\beta - \hat{\mathbb\beta}||$ and the Average Interval Width for the 95\% credible interval. This indicates that our proposed model does not overfit the data. It can be concluded that for cases when dimension reduction is not required the Bayesian GLS model is the preferred model.

\textit{Second Scenario -- the tensor response dimension remains constant while the true dimension of the envelope changes while the variance of the material part is larger than the variance of the immaterial part:} In this scenario, we consider a model where the response is an order-3 tensor with response dimensions $(r_1, r_2, r_3) = (20,20,20)$ and the true envelope dimensions are $(u_1, u_2, u_3) = (5,5,5), (10, 10, 10)$, and $(15, 15, 15)$. For this scenario, we let $\boldsymbol\Omega_{1k}$ to be a $u_k\times u_k$ diagonal matrix with diagonal elements of random uniform $(5,10)$ in descending order. The matrix $\boldsymbol\Omega_{0k}$ is an $(r_k-u_k)\times (r_k-u_k)$ diagonal matrix with diagonal elements of random uniform $(1, 4)$ in descending order.

\begin{scriptsize}
\begin{table}[htp]
\caption{ Average and Standard Deviation of $||\mathbb\beta-\hat{\mathbb\beta}||$ and Average Interval Width (AIW) for 95\% credible interval for an order-3 tensor with response dimensions $(r_1, r_2, r_3) = (20,20,20)$ and the true envelope dimensions are $(u_1, u_2, u_3) = (5,5,5), (10, 10, 10)$, and $(15, 15, 15)$. In this scenario the variance of the material part is larger than the immaterial part.}
\begin{center}
\label{tab2}
\begin{tabular}{|l|l|l|l| l| l| l| l| l| } \hline
 \multicolumn{3}{|c|}{ } &    \multicolumn{2}{c|}{(5,5,5) }       &  \multicolumn{2}{c|}{(10,10,10)}  &  \multicolumn{2}{c|}{(15, 15, 15)}      \\ 
\hline 
 \multicolumn{2}{|c|}{ } &n &	30	&	50	&	30	&	50	&	30	&	50	\\				
\hline															
BTR-OLS&$||\mathbb\beta-\hat{\mathbb\beta}||$	&	mean	&0.782	&	0.526	&	1.773	&	1.280	&	3.498	&	2.570	\\
&	&	sd	&0.157	&	0.058	&	0.274	&	0.170	&	0.608	&	0.246	\\
&	AIW&mean	&35.981	&	36.672	&	55.214	&	56.360	&	78.268	&	79.614	\\
&	&sd	&1.667	&	1.400	&	2.664	&	2.889	&	3.650	&	3.003	\\
\hline															
BTR-GLS&$||\mathbb\beta-\hat{\mathbb\beta}||$	&	mean	&0.563	&	0.297	&	1.262	&	0.747	&	2.448	&	1.484	\\
&	&	sd	&0.147	&	0.046	&	0.247	&	0.122	&	0.565	&	0.205	\\
&	AIW&mean	&10.794	&	10.996	&	17.708	&	18.114	&	22.814	&	23.523	\\
&	&sd	&0.593	&	0.465	&	1.029	&	1.079	&	1.413	&	1.373	\\
\hline															
BTR-ENV	&$||\mathbb\beta-\hat{\mathbb\beta}||$&	mean	&0.019	&	0.016	&	0.138	&	0.132	&	0.740	&	0.525	\\
	&	& sd	&0.002	&	0.002	&	0.005	&	0.005	&	0.148	&	0.046	\\
&	AIW&mean	&0.129	&	0.145	&	0.884	&	0.685	&	3.948	&	3.244	\\
&	&sd	&0.026	&	0.026	&	0.098	&	0.060	&	0.480	&	0.219	\\
\hline
\end{tabular}
\end{center}			
\end{table}
\end{scriptsize}

Table \ref{tab2} shows the results of the simulation for cases where the structural dimension of the envelope is lower than the original tensor response. In this scenario, the dimension of tensor response is fixed and the structural dimension of the envelope changes. In other words, it is the case that the dimension of the material part increases while the dimension of the immaterial part decreases. Furthermore, the variation of the material part is larger than the variation of the immaterial part. Since the computational times for these cases are qualitatively similar to the one in the first scenario, we did not present them. As it can be seen, our proposed model outperforms other models across all different dimensions. 

\textit{Third Scenario -- the tensor response dimension remains constant while the true dimension of the envelope changes while the variance of the material part is less than the variance of the immaterial part:} In this scenario, we consider a model where the response is an order-3 tensor with response dimensions $(r_1, r_2, r_3) = (20,20,20)$ and the true envelope dimensions are $(u_1, u_2, u_3) = (5,5,5), (10, 10, 10)$, and $(15, 15, 15)$. For this scenario, we let $\boldsymbol\Omega_{1k}$ to be a $u_k\times u_k$ diagonal matrix with diagonal elements of random uniform $(1, 5)$ in descending order. The matrix $\boldsymbol\Omega_{0k}$ is an $(r_k-u_k)\times (r_k-u_k)$ diagonal matrix with diagonal elements of random uniform $(15, 20)$ in descending order.

\begin{scriptsize}
\begin{table}[htp]
\caption{ Average and Standard Deviation of $||\mathbb\beta-\hat{\mathbb\beta}||$ and Average Interval Width (AIW) for 95\% credible interval for an order-3 tensor with response dimensions $(r_1, r_2, r_3) = (20,20,20)$ and the true envelope dimensions are $(u_1, u_2, u_3) = (5,5,5), (10, 10, 10)$, and $(15, 15, 15)$. In this scenario, the variation of the immaterial part is larger that the material part. }
\begin{center}
\label{tab3}
\begin{tabular}{|l|l|l|l| l| l| l| l| l| } \hline
 \multicolumn{3}{|c|}{ } &    \multicolumn{2}{c|}{(5,5,5) }       &  \multicolumn{2}{c|}{(10,10,10)}  &  \multicolumn{2}{c|}{(15, 15, 15)}      \\ 
\hline 
 \multicolumn{2}{|c|}{ } &n &	30	&	50	&	30	&	50	&	30	&	50	\\				
\hline															
BTR-OLS&$||\mathbb\beta-\hat{\mathbb\beta}||$	&	mean	&	185.638	&	129.782	&	65.359	&	45.226	&	15.889	&	11.763	\\
	& &	sd	&	27.756	&	12.271	&	10.117	&	4.088	&	2.244	&	1.352	\\
& AIW &	mean	&	671.673	&	605.329	&	340.583	&	341.662	&	167.773	&	171.526	\\
& &	sd	&	49.654	&	28.414	&	10.202	&	10.339	&	5.465	&	6.332	\\
\hline															
BTR-GLS&$||\mathbb\beta-\hat{\mathbb\beta}||$	&	mean	&	108.833	&	69.333	&	48.266	&	28.334	&	11.678	&	7.343	\\
&	&	sd	&	22.935	&	9.305	&	9.807	&	3.577	&	2.082	&	1.081	\\
&AIW &	mean	&	203.617	&	205.472	&	155.004	&	162.711	&	74.467	&	74.841	\\
& &	sd	&	18.507	&	17.636	&	9.343	&	12.180	&	3.629	&	3.941	\\
\hline															
BTR-ENV&$||\mathbb\beta-\hat{\mathbb\beta}||$	&	mean	&	4.159	&	2.438	&	1.382	&	0.737	&	0.467	&	0.299	\\
&	&	sd	&	1.005	&	0.456	&	3.413	&	2.324	&	0.100	&	0.066	\\
&AIW&	mean	&	1.472	&	1.118	&	3.224	&	2.416	&	5.193	&	4.192	\\
&	& sd	&	0.146	&	0.108	&	1.180	&	0.818	&	0.564	&	0.525	\\
\hline
\end{tabular}
\end{center}			
\end{table}
\end{scriptsize}

Table \ref{tab3} shows the results of the simulation for cases where the structural dimension of the envelope is lower than the original tensor response, similar to the second scenario. However, in this scenario, the dimension of tensor response is fixed and the dimension of the envelope increases. In other words, it is the case that the dimension of the material part increases while the dimension of the immaterial part decreases. Since the computational times for these cases are qualitatively similar to the one in the first scenario, we did not present them.  As it can be seen, our proposed model outperforms other models across all different dimensions. Furthermore, as the dimension of the material part of the tensor increases, the estimation accuracy for the Bayesian OLS and Bayesian GLS improves.

\textit{Fourth Scenario -- the dimension of the tensor response changes while the true dimension of the envelope is fixed:} In this scenario, we consider a model where the response is an order-3 tensor with response dimensions $(r_1, r_2, r_3) = (20,20,20), (50, 50, 50)$, and $(100,100,100)$ and the true envelope dimensions are $(u_1, u_2, u_3) = (10, 10, 10)$. For this scenario, we let $\boldsymbol\Omega_{1k}$ to be a $u_k\times u_k$ diagonal matrix with diagonal elements of random uniform $(1, 5)$ in descending order. The matrix $\boldsymbol\Omega_{0k}$ is an $(r_k-u_k)\times (r_k-u_k)$ diagonal matrix with diagonal elements of random uniform $(15, 20)$ in descending order. 

\begin{scriptsize}
\begin{table}[htp]
\caption{ Average and Standard Deviation of $||\mathbb\beta-\hat{\mathbb\beta}||$ and Average Interval Width (AIW) for 95\% credible interval for an order-3 tensor with response dimensions $(r_1, r_2, r_3) = (20,20,20), (50, 50, 50)$, and $(100,100,100)$ and the true envelope dimensions are $(u_1, u_2, u_3) = (10, 10, 10)$.}
\begin{center}
\label{tab4}
\begin{tabular}{|l|l|l|l| l| l| l| l| l| } \hline
 \multicolumn{3}{|c|}{ } &    \multicolumn{2}{c|}{(20,20,20) }       &  \multicolumn{2}{c|}{(50,50,50)}  &  \multicolumn{2}{c|}{(100, 100, 100)}      \\ 
\hline 
 \multicolumn{2}{|c|}{ } 	&n &	30	&	50	&	30	&	50	&	30	&	50	\\				
\hline															
BTR-OLS&	$||\mathbb\beta-\hat{\mathbb\beta}||$ &	mean	&	65.359	&	45.226	&	281.081	&	224.953	&	546.627	&	487.265	\\
	&	& sd	&	10.117	&	4.088	&	31.363	&	9.520	&	30.135	&	13.907	\\
&	AIW & mean	&	340.583	&	341.662	&	926.564	&	920.058	&	1438.921	&	1458.425	\\
&	&sd	&	10.202	&	10.339	&	16.336	&	13.613	&	13.584	&	11.572	\\
\hline															
BTR-GLS&	$||\mathbb\beta-\hat{\mathbb\beta}||$ &	mean	&	48.266	&	28.334	&	128.991	&	75.353	&	166.570	&	96.867	\\
		& &	sd	&	9.807	&	3.577	&	29.846	&	9.327	&	29.190	&	12.880	\\
&	AIW& mean	&	155.004	&	162.711	&	209.495	&	212.579	&	193.128	&	197.634	\\
&	&sd	&	9.343	&	12.180	&	14.362	&	9.874	&	6.679	&	6.368	\\
\hline															
BTR-ENV	&	$||\mathbb\beta-\hat{\mathbb\beta}||$ &	mean	&	1.382	&	0.737	&	2.109	&	1.184	&	0.279	&	0.159	\\
	& &	sd	&	3.413	&	2.324	&	0.527	&	0.179	&	0.055	&	0.023	\\
&	AIW&mean	&	3.224	&	2.416	&	0.850	&	0.608	&	0.113	&	0.079	\\
&	&sd	&	1.180	&	0.818	&	0.126	&	0.066	&	0.013	&	0.009	\\
\hline
\end{tabular}
\end{center}			
\end{table}
\end{scriptsize}

Table \ref{tab4} shows the results of the simulation for cases where the structural dimension of the envelope is lower than the original tensor response. Furthermore, in this scenario, we kept the dimension of the envelope constant and increased the dimension of the tensor response. In other words, it is the case that the dimension of the material part is fixed and the dimension of the immaterial part increases. Since the computational times for these cases are qualitatively similar to the one in the first scenario, we did not present them. For this scenario, based on the average and the standard error of $||\mathbb\beta - \hat{\mathbb\beta}||$ and the Average Interval Width for the 95\% credible interval, our proposed model outperforms other models in all different dimensions. Furthermore, as the dimension of the tensor increases, for the fixed envelope dimension, the results of the Bayesian OLS and Bayesian GLS become worse. This is an expected result since as the dimension of the immaterial subspace increases it adds to the variance of the error. On the other hand, since our proposed model only concentrates on the material part the results are not affected by the increased variance of the immaterial part. Furthermore, as discussed by \cite{cook2010envelope}, the envelope method provides more gains in efficiency when the variance of the immaterial part is much larger than the variance of the material part. Thus, the results of this scenario indicate that when the structural dimension of the envelope is smaller than the dimension of the tensor, our proposed model is the preferred model.

From these simulations, it can be seen that although our proposed method is slower than the Bayesian OLS and Bayesian GLS approaches, the gains in estimation accuracy outweigh the computational deficiency. Thus, we would recommend using our proposed method over the other two approaches.


\section{Real data}
\label{real}
In this section, the data set presented in Section \ref{summary} is analyzed. It is assumed that the brain fMRI data follows the model (\ref{model1}). The fMRI is modeled as a tensor response and contains $145\times 174 \times145$ voxels. After removing the outside brain areas on the image, the size of each image was reduced to $124\times 161\times 126$. Then, we downsized the images to $41\times54\times 42$ which is our three-dimensional tensor response. We chose six covariates related to subject's alcohol use: alcoholic status, the total number of drinks consumed in the past 7 days, and the number of days drank in the last week, both questions were asked on the subject's last day of the HCP study visit, DSM4 criteria for Alcohol Abuse score, drinks consumed per drinking day in the past 12 months and frequency of any alcohol use in the past 12 months. 
Therefore, the independent data is a $178\times 6$ matrix. 
We used Mean Squared Error (MSE), Mean Squared Prediction Error (MSPE), Average Interval Width (AIW) for 95\% credible interval for Tensor response, and Computational Time in second (CT) to compared the performance of the models. To calculate MSE, for $i = 1,\ldots, n$, we let $\mathbbm{Y}_i$ to be the true value of the responses and $\hat{\mathbbm{Y}}_i$ be the estimated value, then
\[
MSE = \frac{1}{n}\sum_{i=1}^{n}  ||\mathbbm{Y}_i-\hat{\mathbbm{Y}}_i||.
\]
The MSPE is calculated by setting aside 30\% of the data as the test data set and using the remaining 70\% of the data to estimate the regression coefficients, $\boldsymbol\beta$. Then, we used the estimated coefficients to predict the test data and calculated MSPE by 
\[
MSPE = \frac{1}{n_{t}}\sum_{i=1}^{n}  ||\mathbbm{Y}_i-\hat{\mathbbm{Y}}_{(-i)}||. 
\]
where $\mathbbm{Y}_i$ and $\hat{\mathbbm{Y}}_{(-i)}$ are the observed and the predicted response of the i-th image in the test data, respectively, and $n_t$ denotes the sample size of the test data. We have done this procedure 50 times and reported the average of them as MSPE. The structural dimension of the envelope, $u_k$, for $k=1,2,3$, for our proposed method is $u_1 =10$, $u_2 = 10$ and $u_3=2$. To choose the structural dimension of the envelope, we tuned $u_k$, for $k=1,2,3$, to get the best combination of the MSE, MSPE, and AIW. It is worth mentioning that the overestimation of the structural dimension leads to estimation with a higher variance while its underestimation results in bias in estimation. Table \ref{tab5} summarizes the results of the data analysis.

\begin{table}[htp]
\caption{ Mean Squared Error (MSE), Mean Squared Prediction Error (MSPE), Average Interval Width (AIW) for 95\% credible interval for Tensor response, and Computational Time in second (CT) for different models. }
\begin{center}
\label{tab5}
\begin{tabular}{lllll } \hline
Model	&  MSE &	MSPE&AIW & CT(s)	\\
\hline
BTR-OLS	&0.0751	& 0.0722 (0.0071) &1.2425 &10367.39 	\\
BTR-GLS		&0.0205	&0.0723 (0.0142) &0.8234 & 24119.92 		\\
BTR-ENV		&0.0069	& 0.0371 (0.0006)& 0.0009 & 62279.56	\\
\hline
\end{tabular}
\end{center}			
\end{table}

From Table \ref{tab5}, it can be seen MSE and MSPE select the proposed Bayesian tensor envelope model over the OLS and GLS Bayesian tensor models in terms of the fit of the model and prediction accuracy. Furthermore, the proposed method provides a smaller AIW for the coefficients. The proposed model needs more computational time to execute. However, based on its superiority in terms of fitting the data and prediction accuracy, our proposed Bayesian tensor regression model outperforms the Bayesian OLS and Bayesian GLS models.

\begin{figure}
\begin{center}
    \includegraphics[width=0.6\textwidth]{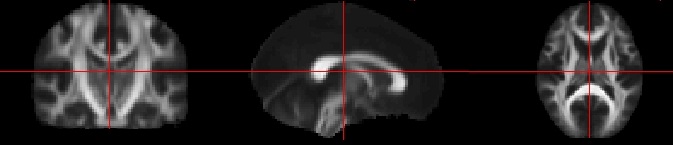}\\
    \includegraphics[width=0.6\textwidth]{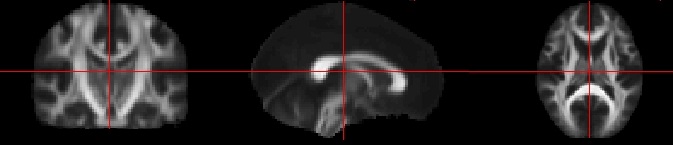}\\
    \includegraphics[width=0.6\textwidth]{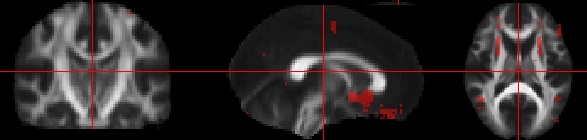}
\caption{Plot of the coefficients and the area of the brain that is activated for Gender. The first row shows the estimated coefficient using Bayesian OLS, second row shows the estimated coefficient using Bayesian GLS, and third row shows the estimated coefficient using proposed Bayesian Tensor Envelope. The red color denotes areas of the brain that the absolute value of the standardized coefficient is above the critical threshold (greater than 16.7261).}
\label{fig1}
\end{center}
\end{figure}

\begin{figure}
\begin{center}
    \includegraphics[width=0.6\textwidth]{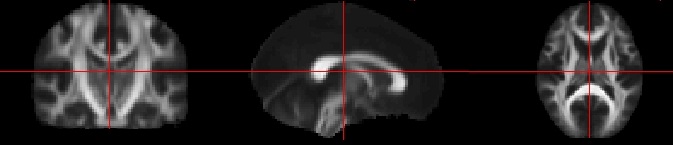}\\
    \includegraphics[width=0.6\textwidth]{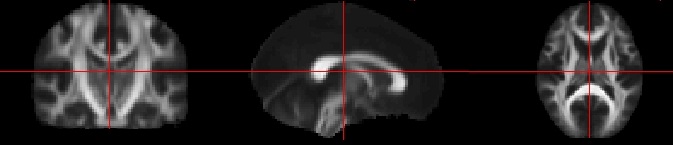}\\
    \includegraphics[width=0.6\textwidth]{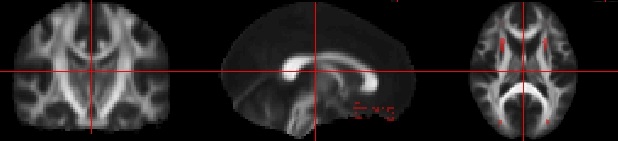}
\caption{Plot of the coefficients and the area of the brain that is activated for drug use. The first row shows the estimated coefficient using Bayesian OLS, second row shows the estimated coefficient using Bayesian GLS, and third row shows the estimated coefficient using proposed Bayesian Tensor Envelope. The red color denotes areas of the brain that the absolute value of the standardized coefficient is above the critical threshold (greater than 16.7261). }
\label{fig2}
\end{center}
\end{figure}

\begin{figure}
\begin{center}
    \includegraphics[width=0.6\textwidth]{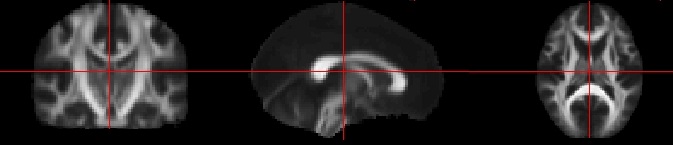}\\
    \includegraphics[width=0.6\textwidth]{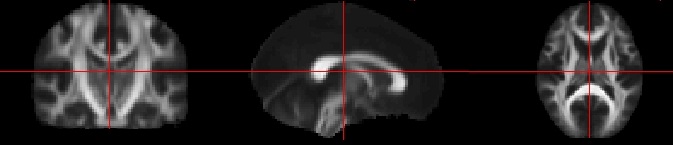}\\
    \includegraphics[width=0.6\textwidth]{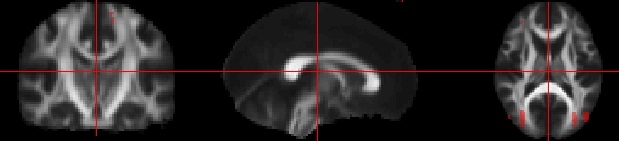}
\caption{Plot of the coefficients and the area of the brain that is activated for past 30 days usage. The first row shows the estimated coefficient using Bayesian OLS, second row shows the estimated coefficient using Bayesian GLS, and third row shows the estimated coefficient using proposed Bayesian Tensor Envelope. The red color denotes areas of the brain that the absolute value of the standardized coefficient is above the critical threshold (greater than 16.7261). }
\label{fig3}
\end{center}
\end{figure}

Figures \ref{fig1}, \ref{fig2}, and \ref{fig3} show the plot of the coefficients and the area of the brain that is activated for Gender, Drug Use, and the past 30 days of usage, respectively. The critical threshold for our study is 16.7261, which was calculated by modifying the critical values formula recommended by \cite{lazzeroni2012cost}. As it can be seen from this figure, using Bayesian OLS and Bayesian GLS approaches, one would fail to reject the null hypothesis that there are areas of the brain that are different between different genders and drug users and controls. However, the proposed Bayesian tensor envelope shows that there are different areas of the brain that are different between the two groups. From these figures, it can be seen that activated areas are located in the Parietal Gyrus, frontal, and internal Gyrus. This area of the brain controls the primary sensory functions, including visual, auditory, and two-dimensional spatial sensations.

\section{Conclusion}
\label{conc}

Alcohol consumption is highly detrimental to brain health. The extensive and prolonged intake of alcohol leads to neuroinflammation, neurodegeneration, and structural alterations in the brain. These effects can result in cognitive impairments, memory loss, decreased executive functioning, and an increased risk of developing alcohol-related brain disorders. The damaging consequences of excessive alcohol consumption on the brain highlight the importance of moderation and responsible drinking to safeguard brain function and overall neurological well-being.
A better understanding of this connection could lead to improved treatments for alcohol abuse. This paper proposes an approach to analyzing data in large dimensions with complex dependency structures. Our proposed model uses a Bayesian regression framework for inference and uses the brain image as a tensor response. The Simulation study shows the model provides an effective inferential tool. Our model is computationally efficient and feasible to implement by using a dimension reduction approach.

\bibliographystyle{plainnat}
\bibliography{papers}

@article{chakraborty2024comprehensive,
  title={A comprehensive Bayesian framework for envelope models},
  author={Chakraborty, Saptarshi and Su, Zhihua},
  journal={Journal of the American Statistical Association},
  volume={119},
  number={547},
  pages={2129--2139},
  year={2024},
  publisher={Taylor \& Francis}
}

@article{giefer2025employing,
  title={Employing tensor regression for analyzing the effect of alcohol on the brain},
  author={Giefer, Clarissa and Ma, Liangsuo and Moradi Rekabdarkolaee, Hossein},
  journal={Arabian Journal of Mathematics},
  pages={1--12},
  year={2025},
  publisher={Springer}
}

@article{ros2018alcohol,
	author = {Ros-Cucurull, Elena and Palma-{\'A}lvarez, Ra{\'u}l Felipe and Cardona-Rubira, Cristina and Garc{\'\i}a-Raboso, Elena and Jacas, Carlos and Grau-L{\'o}pez, Lara and Abad, Alfonso Carlos and Rodr{\'\i}guez-Cintas, Laia and Ros-Montalb{\'a}n, Salvador and Casas, Miguel},
	journal = {Psychiatry Research},
	number = {361--366},
	title = {Alcohol use disorder and cognitive impairment in old age patients: A 6 months follow-up study in an outpatient unit in Barcelona},
	volume = {261},
	year = {2018}}

@article{zahr2014structural,
	author = {Zahr, Natalie M},
	journal = {Handbook of clinical neurology},
	pages = {275--290},
	title = {Structural and microstructural imaging of the brain in alcohol use disorders},
	volume = {125},
	year = {2014}}

@article{crews2014neuroimmune,
	author = {Crews, Fulton T and Vetreno, Ryan P},
	journal = {International review of neurobiology},
	number = {315--357},
	title = {Neuroimmune basis of alcoholic brain damage},
	volume = {118},
	year = {2014}}

@article{sacks2015,
	author = {Sacks, Jeffrey J and Gonzales, Katherine R and Bouchery, Ellen E and Tomedi, Laura E and Brewer, Robert D},
	journal = {American journal of preventive medicine},
	number = {5},
	pages = {e73--e79},
	title = {2010 national and state costs of excessive alcohol consumption},
	volume = {49},
	year = {2015}}

@article{rehm2009global,
	author = {Rehm, J{\"u}rgen and Mathers, Colin and Popova, Svetlana and Thavorncharoensap, Montarat and Teerawattananon, Yot and Patra, Jayadeep},
	journal = {The lancet},
	number = {9682},
	pages = {2223--2233},
	title = {Global burden of disease and injury and economic cost attributable to alcohol use and alcohol-use disorders},
	volume = {373},
	year = {2009}}

@article{deng2021tensor,
	author = {Deng, Kai and Zhang, Xin},
	journal = {Biometrics},
	title = {Tensor envelope mixture model for simultaneous clustering and multiway dimension reduction},
	year = {2021}}

@article{spencer2022parsimonious,
	author = {Spencer, Daniel and Guhaniyogi, Rajarshi and Prado, Raquel},
	journal = {arXiv preprint arXiv:2203.04733},
	title = {Parsimonious Bayesian sparse tensor regression using the Tucker decomposition},
	year = {2022}}

@article{pham2015trace,
	author = {Pham-Gia, T and Thanh, Dinh N and Phong, Duong T},
	journal = {Open Journal of Statistics},
	number = {03},
	pages = {173-190},
	title = {Trace of the Wishart matrix and applications},
	volume = {5},
	year = {2015}}

@article{leebayesian,
	author = {Lee, Minji and Chakraborty, Saptarshi and Su, Zhihua},
	journal = {Statistica Sinica},
	pages = {Accepted},
	title = {A Bayesian approach to envelope quantile regression},
	year = {2022}}

@book{barron1988exponential,
	author = {Barron, Andrew R},
	publisher = {Department of Statistics, University of Illinois Champaign, IL},
	title = {The exponential convergence of posterior probabilities with implications for Bayes estimators of density functions},
	year = {1988}}

@article{bellec2017neuro,
	author = {Bellec, Pierre and Chu, Carlton and Chouinard-Decorte, Francois and Benhajali, Yassine and Margulies, Daniel S and Craddock, R Cameron},
	journal = {Neuroimage},
	pages = {275--286},
	publisher = {Elsevier},
	title = {The Neuro Bureau ADHD-200 Preprocessed repository},
	volume = {144},
	year = {2017}}

@incollection{ch2008remarks,
	author = {Chai, T and Romamoorthi, R. V.},
	booktitle = {Pushing the limits of contemporary statistics: contributions in honor of Jayanta K. Ghosh},
	pages = {170--186},
	publisher = {Institute of Mathematical Statistics},
	title = {Remarks on consistency of posterior distributions},
	year = {2008}}

@article{chen2012sparse,
	author = {Chen, Lisha and Huang, Jianhua Z},
	journal = {Journal of the American Statistical Association},
	number = {500},
	pages = {1533--1545},
	publisher = {Taylor \& Francis Group},
	title = {Sparse reduced-rank regression for simultaneous dimension reduction and variable selection},
	volume = {107},
	year = {2012}}

@article{chun2010sparse,
	author = {Chun, Hyonho and Kele{\c{s}}, S{\"u}nd{\"u}z},
	journal = {Journal of the Royal Statistical Society: Series B (Statistical Methodology)},
	number = {1},
	pages = {3--25},
	publisher = {Wiley Online Library},
	title = {Sparse partial least squares regression for simultaneous dimension reduction and variable selection},
	volume = {72},
	year = {2010}}

@article{cook2010envelope,
	author = {Cook, R Dennis and Li, Bing and Chiaromonte, Francesca},
	journal = {Statistica Sinica},
	pages = {927--960},
	publisher = {JSTOR},
	title = {Envelope models for parsimonious and efficient multivariate linear regression},
	year = {2010}}

@article{goldsmith2014smooth,
	author = {Goldsmith, Jeff and Huang, Lei and Crainiceanu, Ciprian M},
	journal = {Journal of Computational and Graphical Statistics},
	number = {1},
	pages = {46--64},
	publisher = {Taylor \& Francis},
	title = {Smooth scalar-on-image regression via spatial Bayesian variable selection},
	volume = {23},
	year = {2014}}

@article{guhaniyogi2017bayesian,
	author = {Guhaniyogi, Rajarshi and Qamar, Shaan and Dunson, David B},
	journal = {Journal of Machine Learning Research},
	number = {79},
	pages = {1--31},
	title = {Bayesian tensor regression},
	volume = {18},
	year = {2017}}

@article{helland1990partial,
	author = {Helland, Inge S},
	journal = {Scandinavian Journal of Statistics},
	pages = {97--114},
	publisher = {JSTOR},
	title = {Partial least squares regression and statistical models},
	year = {1990}}

@article{jiang2016functional,
	author = {Jiang, Ci-Ren and Aston, John AD and Wang, Jane-Ling},
	journal = {Journal of the American Statistical Association},
	number = {513},
	pages = {1--13},
	publisher = {Taylor \& Francis},
	title = {A functional approach to deconvolve dynamic neuroimaging data},
	volume = {111},
	year = {2016}}

@article{khare2017bayesian,
	author = {Khare, Kshitij and Pal, Subhadip and Su, Zhihua and others},
	journal = {The Annals of Statistics},
	number = {1},
	pages = {196--222},
	publisher = {Institute of Mathematical Statistics},
	title = {A bayesian approach for envelope models},
	volume = {45},
	year = {2017}}

@article{kolda2009tensor,
	author = {Kolda, Tamara G and Bader, Brett W},
	journal = {SIAM review},
	number = {3},
	pages = {455--500},
	publisher = {SIAM},
	title = {Tensor decompositions and applications},
	volume = {51},
	year = {2009}}

@book{lazar2008statistical,
	author = {Lazar, Nicole},
	publisher = {Springer Science \& Business Media},
	title = {The statistical analysis of functional MRI data},
	year = {2008}}

@article{lazzeroni2012cost,
	author = {Lazzeroni, LC and Ray, A},
	journal = {Molecular psychiatry},
	number = {1},
	pages = {108--114},
	publisher = {Nature Publishing Group},
	title = {The cost of large numbers of hypothesis tests on power, effect size and sample size},
	volume = {17},
	year = {2012}}

@article{li2017parsimonious,
	author = {Li, Lexin and Zhang, Xin},
	journal = {Journal of the American Statistical Association},
	pages = {1--16},
	publisher = {Taylor \& Francis},
	title = {Parsimonious tensor response regression},
	year = {2017}}

@article{li2018tucker,
	author = {Li, Xiaoshan and Xu, Da and Zhou, Hua and Li, Lexin},
	journal = {Statistics in Biosciences},
	number = {3},
	pages = {520--545},
	publisher = {Springer},
	title = {Tucker tensor regression and neuroimaging analysis},
	volume = {10},
	year = {2018}}

@article{lindquist2008statistical,
	author = {Lindquist, Martin A},
	journal = {Statistical Science},
	pages = {439--464},
	publisher = {JSTOR},
	title = {The statistical analysis of fMRI data},
	year = {2008}}

@article{manceur2013maximum,
	author = {Manceur, Ameur M and Dutilleul, Pierre},
	journal = {Journal of Computational and Applied Mathematics},
	pages = {37--49},
	publisher = {Elsevier},
	title = {Maximum likelihood estimation for the tensor normal distribution: Algorithm, minimum sample size, and empirical bias and dispersion},
	volume = {239},
	year = {2013}}

@book{ramsay2006functional,
	author = {Ramsay, James O},
	publisher = {Wiley Online Library},
	title = {Functional data analysis},
	year = {2006}}

@article{reiss2010functional,
	author = {Reiss, Philip T and Ogden, R Todd},
	journal = {Biometrics},
	number = {1},
	pages = {61--69},
	publisher = {Wiley Online Library},
	title = {Functional generalized linear models with images as predictors},
	volume = {66},
	year = {2010}}

@article{wang2014regularized,
	author = {Wang, Xuejing and Nan, Bin and Zhu, Ji and Koeppe, Robert},
	journal = {The annals of applied statistics},
	number = {2},
	pages = {1045},
	publisher = {NIH Public Access},
	title = {Regularized 3D functional regression for brain image data via Haar wavelets},
	volume = {8},
	year = {2014}}

@article{yuan2007dimension,
	author = {Yuan, Ming and Ekici, Ali and Lu, Zhaosong and Monteiro, Renato},
	journal = {Journal of the Royal Statistical Society: Series B (Statistical Methodology)},
	number = {3},
	pages = {329--346},
	publisher = {Wiley Online Library},
	title = {Dimension reduction and coefficient estimation in multivariate linear regression},
	volume = {69},
	year = {2007}}

@article{zhang2017tensor,
	author = {Zhang, Xin and Li, Lexin},
	journal = {Technometrics},
	pages = {1--11},
	publisher = {Taylor \& Francis},
	title = {Tensor Envelope Partial Least-Squares Regression},
	year = {2017}}

@article{zhou2013tensor,
	author = {Zhou, Hua and Li, Lexin and Zhu, Hongtu},
	journal = {Journal of the American Statistical Association},
	number = {502},
	pages = {540--552},
	publisher = {Taylor \& Francis Group},
	title = {Tensor regression with applications in neuroimaging data analysis},
	volume = {108},
	year = {2013}}

@article{zhou2014regularized,
	author = {Zhou, Hua and Li, Lexin},
	journal = {Journal of the Royal Statistical Society: Series B (Statistical Methodology)},
	number = {2},
	pages = {463--483},
	publisher = {Wiley Online Library},
	title = {Regularized matrix regression},
	volume = {76},
	year = {2014}}

\newpage
\section{Appendix}

The following notation is used throughout this appendix. We assumed $u_k$, the structural dimension of the envelope for $k=1,\ldots,m$, is known. In addition, we let  $\mathbbm{Y}$ denote the m-order tensor response variable for $n$ observations with dimension $r_1 \times r_2 \times \ldots \times r_m\times n$ and $\textbf{Y}=\mathbbm{Y}_{(m+1)}$ be the $n\times \prod_{i=1}^{m}  r_i $ matrix of the response.  Furthermore, suppose regression coefficient $\mathbb{\beta}$ be the (m+1)-order tensor with dimension $r_1 \times r_2 \times \ldots \times r_m\times p$ and $\boldsymbol\beta=\mathbb{\beta}_{(m+1)}$ be the $p\times \prod_{i=1}^{m}  r_i $ matrix of the regression coefficent. 
\subsection{Consistency of the Posterior Distribution} 
\label{proofcons}
To prove the consistency of posterior distribution, we need to define Kullback-Leiber (KL) support of the prior distribution, $\pi$. Let $K(\theta_0,\theta)=E_{\theta_0} \log(f_{\theta_0}/f_{\theta})$ denotes the KL divergence where  $f(\cdot)$ is the distribution for a given parameter. For $\theta_0$ in the parameter space of $f_\theta$, the set defined as $\{\theta: K(\theta_0,\theta)<\epsilon\}$, for $\epsilon>0$, is called KL neighborhood of $\theta_0$. 

\begin{definition}
\label{def1}
A point $\theta_0$ is in KL support of $\pi$ if for all $\epsilon>0$
\begin{equation*}
\pi(\{\theta: K(\theta_0,\theta)<\epsilon\})>0.
\end{equation*}
where $\pi(\cdot)$ denotes prior distribution. 
\end{definition}

\noindent Adapting the theorem in \cite{ch2008remarks}  and \cite{barron1988exponential} for consistency of the posterior distribution for our proposed Bayesian Tensor Envelope Model, we have the following statement.

\begin{definition}
\label{def2}
Suppose $\mathbb{\eta}^{True}$ is in KL support of the prior distribution and $\underline{\boldsymbol\eta_{n}}$ is the Kulback-Leibler  neighborhood of $\mathbb{\eta}^{True}$. Let $\Pi(\cdot)$ denotes the posterior distribution and $\boldsymbol\eta^{True}=\mathbb{\eta}^{True}_{(m+1)}$. Then, $\Pi_{\boldsymbol\eta^{True}}(\underline{\boldsymbol\eta_{n}^c}|Y,X)$ goes to zero, is equivalent to existence of a sequence of tests $\{\phi_n\}$, $\phi_n$ based on $n$ observations, such that 
\begin{equation}
\label{SAII1}
P_{\boldsymbol\eta^{True}}(\phi_n>0 ~i.o.)=0 ~~and~~inf_{\boldsymbol\eta \in\underline{\boldsymbol\eta_{n}^c} } E_{\boldsymbol\eta} \phi_n\geq 1-c_1 e^{-nc_2},
\end{equation}
where $c_1$ and $c_2$ are constant  greater than zero.
\end{definition}

Therefore, to prove  the posterior distribution is consistent, we must show the following steps
\begin{enumerate}
\item[] Step 1: Find the Kulback-Leibler  neighborhood of $\mathbb{\eta}^{True}$, 
\item[] Step 2: Show the conditions that $\mathbb{\eta}^{True}$ is in Kulback-Leibler support of the priors, 
\item[] Step 3: Find a sequence of tests that satisfies the conditions in equation (\ref{SAII1}).
\end{enumerate}

\noindent\textbf{Step 1:}

\noindent Let $\mathbbm{Y}_i$ be the $i$th observed response, $y_{i}=\mathbbm{Y}_{i(m+1)}$ be the $1\times \prod_{i=1}^{m}r_{i}$ vector of the $i$th observed response, and $\textbf{X}_{i}$ be $i$th observed covariate. For Tensor normal distribution, the KL divergence for each observation is
\begin{footnotesize}
\begin{eqnarray}
&&KL_{i}\left(f(y_{i}|\boldsymbol\eta^{True}),f(y_{i}|\boldsymbol\eta) \right) = \int f(y_{i}|\boldsymbol\eta^{True}) \log\left( \frac{f(y_{i}|\boldsymbol\eta^{True})}{f(y_{i}|\boldsymbol\eta)}\right) dy \cr
&&=  \int f(y_{i}|\boldsymbol\eta^{True}) \bigg\{-\frac{1}{2}\left[(y_{i}\boldsymbol\Gamma_{1}-\textbf{X}_{i}\boldsymbol\eta^{True}) \boldsymbol\Omega_{1}^{-1}(y_{i}\boldsymbol\Gamma_{1}-\textbf{X}_{i}\boldsymbol\eta^{True})^{T}\right] \cr
&&~~~~~~~~~~~~~~+ \frac{1}{2} \left[ (y_{i}\boldsymbol\Gamma_{1}-\textbf{X}_{i}\boldsymbol\eta\pm\textbf{X}_{i}\boldsymbol\eta^{True})\boldsymbol\Omega_{1}^{-1} (y_{i}\boldsymbol\Gamma_{1}-\textbf{X}_{i}\boldsymbol\eta\pm\textbf{X}_{i}\boldsymbol\eta^{True})^{T}\right]\bigg\} dy\cr
&&=  \int f(y_{i}|\boldsymbol\eta^{True}) \left[ (y_{i}\boldsymbol\Gamma_{1}-\textbf{X}_{i}\boldsymbol\eta^{True}) \boldsymbol\Omega_{1}^{-1} (\boldsymbol\eta^{True}-\boldsymbol\eta)^{T}\textbf{X}_{i}^{T} + \frac{1}{2} \textbf{X}_{i}(\boldsymbol\eta^{True}-\boldsymbol\eta)\boldsymbol\Omega_{1}^{-1} (\boldsymbol\eta^{True}-\boldsymbol\eta)^{T}\textbf{X}_{i}^{T} \right] dy \cr
&&= \frac{1}{2} \textbf{X}_{i}(\boldsymbol\eta^{True}-\boldsymbol\eta)\boldsymbol\Omega_{1}^{-1} (\boldsymbol\eta^{True}-\boldsymbol\eta)^{T}\textbf{X}_{i}^{T}.
\end{eqnarray}
\end{footnotesize}
Therefore, the Kulback-Leibler neighborhood for $\boldsymbol\eta^{True}$ is 
\begin{equation*}
\underline{\boldsymbol\eta_{n}} = \left\{ \boldsymbol\eta: \frac{1}{2}\sum_{i=1}^{n} \textbf{X}_{i}(\boldsymbol\eta^{True}-\boldsymbol\eta)\boldsymbol\Omega_{1}^{-1} (\boldsymbol\eta^{True}-\boldsymbol\eta)^{T}\textbf{X}_{i}^{T}< n\epsilon \right\}.
\end{equation*}

\noindent\textbf{Step 2:}

\noindent Based on definition \ref{def1}, $\mathbb{\eta}^{True}$ is in KL support of $\pi$ if for all $\epsilon>0$, we have 
\begin{equation*}
\pi(\underline{\boldsymbol\eta_{n}} ) = \pi\left(\left\{ \boldsymbol\eta: \frac{1}{2}\sum_{i=1}^{n} \textbf{X}_{i}(\boldsymbol\eta^{True}-\boldsymbol\eta)\boldsymbol\Omega_{1}^{-1} (\boldsymbol\eta^{True}-\boldsymbol\eta)^{T}\textbf{X}_{i}^{T}< n\epsilon \right\}\right)>0.
\end{equation*}
Here,
\begin{eqnarray*}
\pi(\underline{\boldsymbol\eta_{n}} )&=&\pi\left(\frac{1}{2}\sum_{i=1}^{n} \textbf{X}_{i}(\boldsymbol\eta^{True}-\boldsymbol\eta)\boldsymbol\Omega_{1}^{-1} (\boldsymbol\eta^{True}-\boldsymbol\eta)^{T}\textbf{X}_{i}^{T}< n\epsilon\right)\cr 
&=& \pi\left( tr\left\{\textbf{X}(\boldsymbol\eta-\boldsymbol\eta^{True}) \boldsymbol\Omega_{1}^{-1} (\boldsymbol\eta -\boldsymbol\eta^{True})^{T}\textbf{X}^{T} \right\} < 2n\epsilon\right).
\end{eqnarray*}
Since for square matrices $\textbf{A}$ and $\textbf{B}$, we have $tr(\textbf{AB})\leq tr(\textbf{A})tr(\textbf{B})$, therefore,
\begin{footnotesize}
\begin{eqnarray*}
\pi\left( tr\left[\textbf{X}(\boldsymbol\eta - \boldsymbol\eta^{True}) \boldsymbol\Omega_{1}^{-1} (\boldsymbol\eta -\boldsymbol\eta^{True})^{T}\textbf{X}^{T} \right] < 2n\epsilon \right)\geq
\pi\left( tr\left[(\boldsymbol\eta - \boldsymbol\eta^{True}) \boldsymbol\Omega_{1}^{-1} (\boldsymbol\eta - \boldsymbol\eta^{True})^{T} \right] < \frac{2n\epsilon}{tr(\textbf{X}^{T}\textbf{X})} \right).
\end{eqnarray*}
\end{footnotesize}
Then, $\eta^{True}$ is in KL support of $\pi$ if for all $\epsilon>0$ 
\begin{eqnarray*}
\pi\left( tr\left[(\boldsymbol\eta - \boldsymbol\eta^{True}) \boldsymbol\Omega_{1}^{-1} (\boldsymbol\eta - \boldsymbol\eta^{True})^{T} \right] < \frac{2n\epsilon}{tr(\textbf{X}^{T}\textbf{X})} \right)>0.
\end{eqnarray*}

\noindent \textbf{Step 3:}

\noindent 
Given $\boldsymbol\Gamma_1$, we have $\boldsymbol\beta = \boldsymbol\Gamma_1\boldsymbol\eta$ and $\boldsymbol\beta^{True} =  \boldsymbol\Gamma_1\boldsymbol\eta^{True}$, then the hypothesis $H_{0}:~ \boldsymbol\beta = \boldsymbol\beta^{True}$ versus $H_{a}:~ \boldsymbol\beta \neq \boldsymbol\beta^{True}$ is equivalent to the hypothesis $H_{0}:~ \boldsymbol\eta = \boldsymbol\eta^{True}$ versus $H_{a}:~ \boldsymbol\eta \neq \boldsymbol\eta^{True}$. This equivalent is due to orthogonality of $\boldsymbol\Gamma_1$. Define a sequence of tests $\{\phi_n\}$ for hypothesis  $H_{0}:~ \boldsymbol\eta = \boldsymbol\eta^{True}$ versus $H_{a}:~ \boldsymbol\eta \neq \boldsymbol\eta^{True}$ with Criteria Region $C_{n} = \left\{ \boldsymbol\eta : \frac{1}{n} \sum_{i=1}^{n} (y_{i}\boldsymbol\Gamma_{1}-\textbf{X}_{i}\boldsymbol\eta^{True}) \boldsymbol\Omega_{1}^{-1}(y_{i}\boldsymbol\Gamma_{1}-\textbf{X}_{i}\boldsymbol\eta^{True})^{T} > \frac{\epsilon}{4} \right\}$. Since
\begin{small}
\[
tr\left[ (Y\boldsymbol\Gamma_{1}-\textbf{X}\boldsymbol\eta^{True}) \boldsymbol\Omega_{1}^{-1}(Y\boldsymbol\Gamma_{1}-\textbf{X}\boldsymbol\eta^{True})^{T} \right] = \sum_{i=1}^{n} (y_{i}\boldsymbol\Gamma_{1}-\textbf{X}_{i}\boldsymbol\eta^{True}) \boldsymbol\Omega_{1}^{-1}(y_{i}\boldsymbol\Gamma_{1}-\textbf{X}_{i}\boldsymbol\eta^{True})^{T}
\]
\end{small}
and under the null hypothesis 
\[
 \sum_{i=1}^{n} (y_{i}\boldsymbol\Gamma_{1}-\textbf{X}_{i}\boldsymbol\eta^{True}) \boldsymbol\Omega_{1}^{-1}(y_{i}\boldsymbol\Gamma_{1}-\textbf{X}_{i}\boldsymbol\eta^{True})^{T} | \boldsymbol\Gamma_{1},\boldsymbol\Omega_{1} \sim \chi_{n\prod_{j=1}^{m} u_{j}}^{2}.
\]
then, for large n, we have
\begin{footnotesize}
\begin{equation}
\label{s1}
E_{\boldsymbol\eta^{True}}(\Phi_{n} | \boldsymbol\Gamma_{1},\boldsymbol\Omega_{1}) = P_{\boldsymbol\eta^{True}}\left(\sum_{i=1}^{n} (y_{i}\boldsymbol\Gamma_{1}-\textbf{X}_{i}\boldsymbol\eta^{True}) \boldsymbol\Omega_{1}^{-1}(y_{i}\boldsymbol\Gamma_{1}-\textbf{X}_{i}\boldsymbol\eta^{True})^{T} > \frac{n\epsilon}{4} | \boldsymbol\Gamma_{1},\boldsymbol\Omega_{1} \right) \leq \exp\left( -\frac{n\epsilon}{16} \right).~~~
\end{equation}
\end{footnotesize}
From Markov inequality, we have $P(x_{n}>x)\leq \frac{E(x_{n})}{x}$. Therefore, 
\begin{eqnarray*}
P_{\boldsymbol\eta^{True}} \left(\Phi_n > \exp\left\{-\frac{b_{1}n}{2}\right\}|\boldsymbol\Gamma_{1},\boldsymbol\Omega_{1}\right) &\leq& E_{\boldsymbol\eta^{True}} \left(\Phi_n|\boldsymbol\Gamma_{1},\boldsymbol\Omega_{1}\right) \exp\left\{\frac{b_{1}n}{2}\right\}\cr
&\leq& \exp\left( -\frac{n\epsilon}{16} \right)\exp\left\{\frac{b_{1}n}{2}\right\},
\end{eqnarray*}
let $b_1=\frac{\epsilon}{16}$ then,
\begin{eqnarray}
P_{\boldsymbol\eta^{True}} \left(\Phi_n > \exp\left\{-\frac{b_{1}n}{2}\right\}\right) &=& E\left[ P_{\boldsymbol\eta^{True}} \left(\Phi_n > \exp\left\{-\frac{b_{1}n}{2}\right\}|\boldsymbol\Gamma_{1},\boldsymbol\Omega_{1}\right) \right]\cr
&\leq&E_{\boldsymbol\Gamma_{1},\boldsymbol\Omega_{1}} \left[\exp\left\{-\frac{n\epsilon}{32}\right\}\right] = \exp\left\{-\frac{n\epsilon}{32}\right\} \rightarrow 0.~~~~~~~~
\end{eqnarray}
This proves the first condition in \ref{def2}. To show the second condition in \ref{def2}, similar to \cite{guhaniyogi2017bayesian}, we have
\begin{small}
\begin{eqnarray}
&\frac{1}{n}& tr\left[ (Y\boldsymbol\Gamma_{1}-\textbf{X}\boldsymbol\eta^{True}) \boldsymbol\Omega_{1}^{-1}(Y\boldsymbol\Gamma_{1}-\textbf{X}\boldsymbol\eta^{True})^{T} \right] \cr
&=& \frac{1}{n} tr\left[ (Y\boldsymbol\Gamma_{1}-\textbf{X}\boldsymbol\eta^{True}\pm \textbf{X}\boldsymbol\eta) \boldsymbol\Omega_{1}^{-1}(Y\boldsymbol\Gamma_{1}-\textbf{X}\boldsymbol\eta^{True}\pm \textbf{X}\boldsymbol\eta)^{T} \right] \cr
&=& \frac{1}{n} tr \left[ (Y\boldsymbol\Gamma_{1}-\textbf{X}\boldsymbol\eta) \boldsymbol\Omega_{1}^{-1}(Y\boldsymbol\Gamma_{1}-\textbf{X}\boldsymbol\eta)^{T} + \textbf{X}(\boldsymbol\eta^{True}-\boldsymbol\eta)\boldsymbol\Omega_{1}^{-1} (\boldsymbol\eta^{True}-\boldsymbol\eta)^{T}\textbf{X}^{T} \right] \cr
&+& \frac{1}{n} tr \left[ (Y\boldsymbol\Gamma_{1}-\textbf{X}\boldsymbol\eta) \boldsymbol\Omega_{1}^{-1}(\boldsymbol\eta^{True}-\boldsymbol\eta)^{T}\textbf{X}^{T} + \textbf{X}(\boldsymbol\eta^{True}-\boldsymbol\eta)\boldsymbol\Omega_{1}^{-1} (Y\boldsymbol\Gamma_{1}-\textbf{X}\boldsymbol\eta)^{T} \right] \cr
&=& \frac{1}{n} tr \left[ (Y\boldsymbol\Gamma_{1}-\textbf{X}\boldsymbol\eta) \boldsymbol\Omega_{1}^{-1}(Y\boldsymbol\Gamma_{1}-\textbf{X}\boldsymbol\eta)^{T} \right] +  \frac{1}{n} tr \left[ \textbf{X}(\boldsymbol\eta^{True}-\boldsymbol\eta)\boldsymbol\Omega_{1}^{-1} (\boldsymbol\eta^{True}-\boldsymbol\eta)^{T}\textbf{X}^{T} \right] \cr
&+& \frac{2}{n} tr \left[ (Y\boldsymbol\Gamma_{1}-\textbf{X}\boldsymbol\eta) \boldsymbol\Omega_{1}^{-1}(\boldsymbol\eta^{True}-\boldsymbol\eta)^{T}\textbf{X}^{T} \right]
\end{eqnarray}
\end{small}
Since,
\[
\frac{2}{n} tr \left[ (Y\boldsymbol\Gamma_{1}-\textbf{X}\boldsymbol\eta) \boldsymbol\Omega_{1}^{-1}(\boldsymbol\eta^{True}-\boldsymbol\eta)^{T}\textbf{X}^{T} \right] | \boldsymbol\Gamma_{1},\boldsymbol\Omega_{1} \sim N\left(0,\frac{8}{n^{2}}\sum_iKL_i\right),
\]
thus, 
\[
\frac{2}{n} tr \left[ (Y\boldsymbol\Gamma_{1}-\textbf{X}\boldsymbol\eta) \boldsymbol\Omega_{1}^{-1}(\boldsymbol\eta^{True}-\boldsymbol\eta)^{T}\textbf{X}^{T} \right] \equiv \sqrt{\frac{8}{n}\sum_iKL_i}\frac{Z}{\sqrt{n}},
\]
where $Z\sim N(0,1)$. Therefore, 
\begin{footnotesize}
\begin{eqnarray*}
&\sup_{\boldsymbol\eta\in  \underline{\boldsymbol\eta_{n}}^{c}}& E_{\boldsymbol\eta}(1-\Phi_{n} | \boldsymbol\Gamma_{1},\boldsymbol\Omega_{1}) = \sup_{\boldsymbol\eta\in  \underline{\boldsymbol\eta_{n}}^{c}} P_\eta\left(\frac{1}{n} tr \left[ (y_{i}\boldsymbol\Gamma_{1}-\textbf{X}\boldsymbol\eta^{True}) \boldsymbol\Omega_{1}^{-1}(y_{i}\boldsymbol\Gamma_{1}-\textbf{X}\boldsymbol\eta^{True})^{T} \right]\leq \frac{\epsilon}{4}  | \boldsymbol\Gamma_{1},\boldsymbol\Omega_{1}\right)\cr
&=& \sup_{\boldsymbol\eta\in  \underline{\boldsymbol\eta_{n}}^{c}} P_\eta\left\{ \left(\frac{1}{n}  tr \left[ (y_{i}\boldsymbol\Gamma_{1}-\textbf{X}\boldsymbol\eta) \boldsymbol\Omega_{1}^{-1}(y_{i}\boldsymbol\Gamma_{1}-\textbf{X}\boldsymbol\eta)^{T} \right] + \frac{2}{n} \sum_iKL_i + \sqrt{\frac{8}{n}\sum_iKL_i}\frac{Z}{\sqrt{n}}\right) \leq \frac{\epsilon}{4}  | \boldsymbol\Gamma_{1},\boldsymbol\Omega_{1}\right\}\cr
&\leq&  \sup_{\boldsymbol\eta\in  \underline{\boldsymbol\eta_{n}}^{c}} P_\eta\left(\left\{ \left|\frac{2}{n}\sum_iKL_i + \sqrt{\frac{8}{n}\sum_iKL_i}\frac{Z}{\sqrt{n}}\right|- \frac{\epsilon}{4} \leq \left|\frac{1}{n}  tr \left[ (y_{i}\boldsymbol\Gamma_{1}-\textbf{X}\boldsymbol\eta) \boldsymbol\Omega_{1}^{-1}(y_{i}\boldsymbol\Gamma_{1}-\textbf{X}\boldsymbol\eta)^{T} \right] \right|\right\} | \boldsymbol\Gamma_{1},\boldsymbol\Omega_{1}\right).
\end{eqnarray*}
\end{footnotesize}
Let $A = \left\{  \left|\frac{2}{n} \sum_iKL_i + \sqrt{\frac{8}{n}\sum_iKL_i}\frac{Z}{\sqrt{n}}\right|- \frac{\epsilon}{4} \leq \left|\frac{1}{n}  tr \left[ (y_{i}\boldsymbol\Gamma_{1}-\textbf{X}\boldsymbol\eta) \boldsymbol\Omega_{1}^{-1}(y_{i}\boldsymbol\Gamma_{1}-\textbf{X}\boldsymbol\eta)^{T} \right] \right| \right\}$ and $\tau_{n} = \left\{\boldsymbol\eta:  \left| \sqrt{\frac{8}{n}\sum_iKL_i}\frac{Z}{\sqrt{n}}\right| \leq \frac{1}{n} \sum_iKL_i  \right\}$, i.e. 
\[
-\frac{1}{n} \sum_iKL_i \leq  \sqrt{\frac{8}{n}\sum_iKL_i}\frac{Z}{\sqrt{n}} \leq \frac{1}{n} \sum_iKL_i,
\]
thus, we have
\begin{small}
\begin{eqnarray}
\label{s2}
&\sup_{\boldsymbol\eta\in  \underline{\boldsymbol\eta_{n}}^{c}}& E_{\boldsymbol\eta}(1-\Phi_{n} | \boldsymbol\Gamma_{1},\boldsymbol\Omega_{1}) \leq \sup_{\boldsymbol\eta\in  \underline{\boldsymbol\eta_{n}}^{c}} P_{\boldsymbol\eta}(A\cap \tau_{n} | \boldsymbol\Gamma_{1},\boldsymbol\Omega_{1}) +\sup_{\boldsymbol\eta\in  \underline{\boldsymbol\eta_{n}}^{c}} P_{\boldsymbol\eta}(A\cap \tau_{n}^{c} | \boldsymbol\Gamma_{1},\boldsymbol\Omega_{1})\cr 
&\leq & \sup_{\boldsymbol\eta\in  \underline{\boldsymbol\eta_{n}}^{c}} P_{\boldsymbol\eta}(A\cap \tau_{n} | \boldsymbol\Gamma_{1},\boldsymbol\Omega_{1}) +\sup_{\boldsymbol\eta\in  \underline{\boldsymbol\eta_{n}}^{c}} P_{\boldsymbol\eta}(\tau_{n}^{c} | \boldsymbol\Gamma_{1},\boldsymbol\Omega_{1})\cr 
&\leq & \sup_{\boldsymbol\eta\in  \underline{\boldsymbol\eta_{n}}^{c}} P_{\boldsymbol\eta}\left\{\left( \frac{1}{n}\sum_iKL_i - \frac{\epsilon}{4} \leq \left|\frac{1}{n}  tr \left[ (y_{i}\boldsymbol\Gamma_{1}-\textbf{X}\boldsymbol\eta) \boldsymbol\Omega_{1}^{-1}(y_{i}\boldsymbol\Gamma_{1}-\textbf{X}\boldsymbol\eta)^{T} \right] \right| \right) | \boldsymbol\Gamma_{1},\boldsymbol\Omega_{1}\right\} \cr
&+&  P_{\boldsymbol\eta}\left\{ \left(|z|\geq \frac{\sqrt{n\epsilon}}{\sqrt{8}}\right) | \boldsymbol\Gamma_{1},\boldsymbol\Omega_{1}\right\}\cr
&\leq & \exp\left(-\frac{3n\epsilon}{16} \right)+ \exp\left(-\frac{n\epsilon}{32} \right) \leq 2 \exp\left(-\frac{n\epsilon}{32} \right)
\end{eqnarray}
\end{small}
Since
\[
\sup E_{\boldsymbol\eta} (A) = \sup E\left[E_\eta(A| \boldsymbol\Gamma_{1},\boldsymbol\Omega_{1})\right]
\leq E\left[ \sup E_{\boldsymbol\eta} (A| \boldsymbol\Gamma_{1},\boldsymbol\Omega_{1}) \right]
\]
thus,
\begin{eqnarray}
\sup_{\boldsymbol\eta\in  \underline{\boldsymbol\eta_{n}}^{c}}E\left( E_{\boldsymbol\eta} \left\{ 1-\Phi_{n}\right\} |\boldsymbol\Gamma_{1},\boldsymbol\Omega_{1}  \right) &\leq& 2 \exp\left(-\frac{n\epsilon}{32} \right),
\end{eqnarray}
which means,
\begin{eqnarray}
1-\inf_{\boldsymbol\eta\in  \underline{\boldsymbol\eta_{n}}^{c}}E\left(  \Phi_{n}   \right) &\leq& 2 \exp\left(-\frac{n\epsilon}{32} \right), 
\end{eqnarray}
therefore, 
\begin{eqnarray}
\inf_{\boldsymbol\eta\in  \underline{\boldsymbol\eta_{n}}^{c}}E\left(  \Phi_{n}   \right) &>&1- 2 \exp\left(-\frac{n\epsilon}{32} \right).
\end{eqnarray}
This proves the second condition in \ref{def2}. From steps 1, 2, and 3, we can see the posterior, i.e. $\Pi_{\boldsymbol\eta^{True}}(\underline{\boldsymbol\eta_{n}^c}|Y,X)$, goes to zero. This completes the proof.

\subsection{Derivation of the posterior for different models}
In this section, we provide the derivation of the posterior distribution for the parameters of different models. The likelihood function form model (\ref{model1}) is 
\begin{equation}
\label{likeap1}
L(\textbf{Y}|\boldsymbol\beta,\boldsymbol\Sigma,\textbf{X}) = |\boldsymbol\Sigma|^{-\frac{n}{2}} \exp \left\{-\frac{1}{2} tr\left[(\textbf{Y}-\textbf{X}\boldsymbol\beta)\boldsymbol\Sigma^{-1}(\textbf{Y}-\textbf{X}\boldsymbol\beta)^{T}\right] \right\}.
\end{equation}
where $\boldsymbol\Sigma=\Sigma_m\otimes\ldots \otimes\Sigma_1$. 
\subsubsection{Posterior for Bayesian OLS}
\label{olsp}
If we let $\Sigma_{k} = \sigma_k^{2}\textbf{I}_{k}$ where $\textbf{I}_{k}$ is an identity matrix with dimension $r_k$, then the likelihood function (\ref{likeap1}) can be rewritten as 
\begin{small}
\begin{eqnarray*}
L(\textbf{Y}|\boldsymbol\beta, \sigma_1^{2},\cdots, \sigma_m^{2},\textbf{X})\propto \left(\prod_{i=1}^{m} \sigma_{i}^{2} \right)^{\frac{-n\prod_{j=1}^{m} r_j}{2}} \exp \left\{ -\frac{1}{2} tr\left[\left(\prod_{i=1}^{m} \sigma_{i}^{2} \right)^{-1} (\textbf{Y}-\textbf{X}\boldsymbol\beta)(\textbf{Y}-\textbf{X}\boldsymbol\beta)^{T}\right] \right\}.
\end{eqnarray*}
\end{small}
For Bayesian OLS, we assumed $\pi(\mathbb{\beta}) \sim TN(\mathbbm{B}, \sigma_{1}^{2} \textbf{I}_1, \ldots, \sigma_{m}^{2}\textbf{I}_m, \textbf{I}_p)$ where $\mathbbm{B}$ be the tensor mean for the prior of the regression coefficient. Then the posterior for the tensor coefficients is 
\[
\Pi(\mathbb{\beta}|\textbf{Y},\sigma_1^{2},\cdots, \sigma_m^{2}, \textbf{X}) \sim TN \left(\mathbb{\mu},\sigma_{1}^{2} \textbf{I}_1, \ldots, \sigma_{m}^{2}\textbf{I}_m,\textbf{E}\right).
\]
where $\mathbb{\mu}=(\mathbbm{Y}\times_{(m+1)}\textbf{X}^T + \mathbbm{B})\times_{(p+1)}(\textbf{X}^{T} \textbf{X} + \textbf{I}_p )^{-1}$ and $\textbf{E}=(\textbf{X}^{T} \textbf{X} + \textbf{I}_p )^{-1}$. Furthermore, for $k=1,\ldots,m$, let $\sigma_{k}^{2}$ follow an inverse-Gamma distribution with positive parameters $\alpha_k$ and $b_k$ i.e. 
\[
\pi(\sigma_{k}^{2}) = (\sigma_{k}^{2})^{-\alpha_k -1} \exp\left\{-\frac{b_{k}}{\sigma_{k}^{2}} \right\}.
\]
Then, for $k=1,\ldots,m$, the posterior distribution for $\sigma_{k}^{2}$ is
\begin{eqnarray*}
\Pi(\sigma_{k}^{2}|\textbf{Y},\textbf{X}) &\propto &  (\sigma_{k}^{2})^{\frac{-n\prod_{j=1}^{m} r_j}{2}-\alpha_k -1} \exp\left\{-\frac{1}{\sigma_{k}^{2}} \left[b_{k}+\frac{1}{2} tr\left[\left(\prod_{i\neq k =1}^{m} \sigma_{i}^{2} \right)^{-1} \textbf{A}\right] \right] \right\} \\
&\Rightarrow&\\
\Pi(\sigma_{k}^{2}|\textbf{Y},\textbf{X}) &\sim& IG\left[n\frac{\prod_{j=1}^{m} r_j}{2}+\alpha_k, b_k +\frac{1}{2} tr\left[\left(\prod_{i\neq k =1}^{m} \sigma_{i}^{2} \right)^{-1} \textbf{A} \right]\right],
\end{eqnarray*}
where $\textbf{A} =  \textbf{Y}^{T}\textbf{Y} - \mathbb{\mu}_{(m+1)}^{T}\textbf{E}^{-1}\mathbb{\mu}_{(m+1)} + \mathbbm{B}_{(m+1)}^{T}\mathbbm{B}_{(m+1)}$.

\subsubsection{Posterior for Bayesian GLS}
\label{glsp}

For Bayesian GLS, we assumed $\boldsymbol\Sigma=\Sigma_m\otimes\ldots \otimes\Sigma_1$ and $\pi(\mathbb{\beta}) \sim TN(\mathbbm{B}, \Sigma_1, \ldots, \Sigma_m, \textbf{I}_p)$. 
\begin{footnotesize}
\begin{eqnarray*}
\pi(\boldsymbol\beta|\boldsymbol\Sigma) &=& |\boldsymbol\Sigma|^{-\frac{p}{2}} \exp \left\{-\frac{1}{2} tr\left[\boldsymbol\Sigma^{-1}(\boldsymbol\beta-\mathbbm{B}_{(m+1)})^{T}(\boldsymbol\beta-\mathbbm{B}_{(m+1)})\right] \right\} \\
&\Rightarrow& \\
\Pi(\boldsymbol\beta|\textbf{Y},\boldsymbol\Sigma, \textbf{X}) &\propto&  \exp \left\{-\frac{1}{2} tr\left[\boldsymbol\Sigma^{-1}(\textbf{Y}-\textbf{X}\boldsymbol\beta)^{T}(\textbf{Y}-\textbf{X}\boldsymbol\beta)  +\boldsymbol\Sigma^{-1}(\boldsymbol\beta-\mathbbm{B}_{(m+1)})^{T}(\boldsymbol\beta-\mathbbm{B}_{(m+1)})\right] \right\}\\
&=& \exp \Bigg\{-\frac{1}{2} tr\bigg[\boldsymbol\Sigma^{-1}\bigg(\textbf{Y}^{T}\textbf{Y}- \boldsymbol\beta^{T}\textbf{X}^{T}\textbf{Y} - \textbf{Y}^{T} \textbf{X}\boldsymbol\beta \cr
&&~~~~~~~+ \boldsymbol\beta^{T}\textbf{X}^{T} \textbf{X}\boldsymbol\beta+ \boldsymbol\beta^{T}\boldsymbol\beta - \boldsymbol\beta^{T}\mathbbm{B} _{(m+1)}- \mathbbm{B}_{(m+1)}^{T}\boldsymbol\beta + \mathbbm{B}_{(m+1)}^{T}\mathbbm{B}_{(m+1)}  \bigg) \bigg] \Bigg\} \cr 
&\propto&  \exp \left\{-\frac{1}{2} tr\left[\boldsymbol\Sigma^{-1}\left(  \boldsymbol\beta^{T}(\textbf{X}^{T} \textbf{X} + \textbf{I}_p )\boldsymbol\beta -  \boldsymbol\beta^{T}(\textbf{X}^{T}\textbf{Y} + \mathbbm{B}_{(m+1)} ) -  (\textbf{Y}^{T}\textbf{X} + \mathbbm{B}_{(m+1)})^{T} \boldsymbol\beta \right) \right] \right\} \\
&\Rightarrow& \\
\Pi(\mathbb{\beta}|\mathbbm{Y},\boldsymbol\Sigma, \textbf{X}) &\sim& TN \left((\mathbbm{Y} \times_{(m+1)} \textbf{X}^{T}+ \mathbbm{B})\times_{(p+1)}(\textbf{X}^{T} \textbf{X} + \textbf{I}_p )^{-1}, \Sigma_{1},\ldots,\Sigma_{m}, (\textbf{X}^{T} \textbf{X} + \textbf{I}_p )^{-1}\right).
\end{eqnarray*}
\end{footnotesize}
Furthermore, for $k=1,\ldots,m$, let $\Sigma_{k}$ follow an inverse-Wishart distribution with parameters $\Psi_k$ and $\nu_k$ i.e. 
\[
\pi(\Sigma_k) \sim IW (\Psi_k,\nu_k) = |\Sigma_{k}|^{-\frac{\nu_k + r_k +1 }{2}} \exp\{-\frac{1}{2} tr(\Sigma_k^{-1}\Psi_k) \}.
\]
Then, for $k=1,\ldots,m$, the posterior distribution for $\Sigma_{k}^{2}$ is
\begin{eqnarray*}
\Pi(\Sigma_{k}|\textbf{Y}, \textbf{X}) &\propto& |\Sigma_{k}|^{-\frac{\nu_k + r_k +1 }{2}} \exp\{-\frac{1}{2} tr(\Sigma_k^{-1}\Psi_k) \times |\boldsymbol\Sigma|^{-\frac{n}{2}} \exp\left\{-\frac{1}{2} tr(\boldsymbol\Sigma^{-1}\textbf{A}) \right\} \\
&=&  |\Sigma_{k}|^{-\frac{\nu_k + r_k +1 + n\prod_{i\neq k =1}^{m} r_{i} }{2}} \exp \left\{ -\frac{1}{2} tr(\Sigma_k^{-1}\Psi_k) -\frac{1}{2}  tr[\tilde{D}_{\textbf{C}_{k}}\Sigma_{k}^{-1}] \right\}\\
&=&  |\Sigma_{k}|^{-\frac{\nu_k + r_k +1 + n\prod_{i\neq k =1}^{m} r_{i} }{2}} \exp \left\{ -\frac{1}{2} tr[\Psi_k+\tilde{D}_{\textbf{C}_{k}}]\Sigma_k^{-1} \right\}\\
&\Rightarrow&\\
\Pi(\Sigma_{k}|\textbf{Y}, \textbf{X})  &\sim& IW(\nu_k + n\prod_{i\neq k =1}^{m} r_{i}, \Psi_k+\tilde{D}_{\textbf{C}_{k}}).
\end{eqnarray*}
In second equation,
\[
tr(\textbf{A}\boldsymbol\Sigma^{-1}) = tr(\textbf{C}_{k}\textbf{A}\textbf{C}_{k}^{T}\textbf{C}_{k}\boldsymbol\Sigma^{-1}\textbf{C}_{k}^{T}) = tr[\textbf{C}_{k}\textbf{A}\textbf{C}_{k}^{T}(\boldsymbol\Sigma^{k\dagger}\otimes\boldsymbol\Sigma_{k}^{\dagger})] = tr[\tilde{D}_{\textbf{C}_{k}}\boldsymbol\Sigma_{k}^{-1}].
\]
Where $\textbf{C}_k=I_{r_m\times \cdots\times r_{k+1}}\otimes K_{r_{k-1}\times\cdots\times r_1,r_k}$, $K_{m,n}$ is a $mn\times mn$ commutation matrix and $\boldsymbol{\Sigma}^{k\dagger}=\boldsymbol{\Sigma}_{m}^{k\dagger}\otimes\cdots\otimes \boldsymbol{\Sigma}_{k+1}^{k\dagger}\otimes \boldsymbol{\Sigma}_{k-1}^{k\dagger}\otimes\cdots\otimes \boldsymbol{\Sigma}_{1}^{k\dagger}=[\sigma^{k\dagger}_{ij}]$. Let $D_{\textbf{C}_{k}}=\textbf{C}_k\textbf{A}\textbf{C}_k^{T}$, if we partition $D_{\textbf{C}_{k}}$ to $r_k \times r_k$ matrices of $D_{ij}$ then $\tilde{D}_{\textbf{C}_{k}}=\sum_{i}\sum_{j} \sigma^{k\dagger}_{ij}D_{ji}$.  Here, $\textbf{A} =  \textbf{Y}^{T}\textbf{Y} - \mathbb{\mu}_{(m+1)}^{T}\textbf{E}^{-1}\mathbb{\mu}_{(m+1)} + \mathbbm{B}_{(m+1)}^{T}\mathbbm{B}_{(m+1)}$ where $\mathbb{\mu}=(\mathbbm{Y}\times_{(m+1)}\textbf{X}^T + \mathbbm{B})\times_{(p+1)}(\textbf{X}^{T} \textbf{X} + \textbf{I}_p )^{-1}$ and $\textbf{E}=(\textbf{X}^{T} \textbf{X} + \textbf{I}_p )^{-1}$.
\subsubsection{Posterior for Proposed Bayesian Tensor Envelope Regression}
\label{tenp}
In this section, we find the full conditional distribution of each parameters. It should be noted that we have the following properties. For the regression coefficients $\mathbb{\beta}$, we assume
\begin{equation}
\mathbb{\beta} = [\mathbb{\eta}; \boldsymbol\Gamma_{11}, \cdots, \boldsymbol\Gamma_{1m},\textbf{I}_{p}]
\end{equation}
let $\boldsymbol\eta=\mathbb{\eta_{(m+1)}}$ then, we have
\begin{equation}
\boldsymbol\beta=I_p\boldsymbol\eta(\boldsymbol\Gamma_{m1}\otimes \cdots \otimes \boldsymbol\Gamma_{11})^{T}=\boldsymbol\eta(\boldsymbol\Gamma_{m1}\otimes \cdots \otimes \boldsymbol\Gamma_{11})^{T}=\boldsymbol\eta\boldsymbol\Gamma_{1}^T.
\end{equation}
The covariance matrix $\boldsymbol\Sigma$ can be written as
\begin{equation}
\boldsymbol\Sigma=\Sigma_m\otimes\cdots\otimes\Sigma_1.
\end{equation}
 Following envelope (Cook et al., 2010), for $k=1,\ldots, m$, the covariance matrix can be decompose as 
\begin{equation}
\Sigma_k= V_{0k}+V_{1k}=\Gamma_{0k}\Omega^{-1}_{0k}\Gamma_{0k}+\Gamma_{1k}\Omega^{-1}_{1k}\Gamma_{1k}
\end{equation}
where $V_{0k}$ and $V_{1k}$ denotes the covariance of the immaterial and material part of the response, respectively. Therefore, the determinant of the covariance function is 
\begin{eqnarray}
|\boldsymbol\Sigma|&=&|\Sigma_m|^{\prod_{k=1}^{m-1}r_{k}}\cdots|\Sigma_1|^{\prod_{k=2}^{m}r_{k}}\cr
&=&(|\Omega_{m1}||\Omega_{m0}|)^{\prod_{k=1}^{m-1}r_{k}}\cdots(|\Omega_{11}||\Omega_{10}|)^{\prod_{k=2}^{m}r_{k}}\cr
&=&\prod_{k=1}^{m}(|\Omega_{k1}||\Omega_{k0}|)^{\prod_{i=1,i\neq k}^{m}r_{i}}=\prod_{k=1}^{m}\left(\prod_{j=1}^{u_k}|\omega_{k1,j}|\prod_{j'=1}^{r_k-u_k}|\omega_{k0,j'}|\right)^{\prod_{i=1,i\neq k}^{m}r_{i}}
\end{eqnarray}
and the inverse of the covariance matrix is
\begin{eqnarray}
\boldsymbol\Sigma^{-1}=\Sigma^{-1}_m\otimes\cdots\otimes\Sigma^{-1}_1=(V^\dagger_{m1}+V^\dagger_{m0})\otimes\cdots\otimes(V^\dagger_{11}+V^\dagger_{10})=\boldsymbol{V}^\dagger_1+\boldsymbol{V}^\dagger_0.
\end{eqnarray}
To find the posteriors, it should be noted that the likelihood function of $\textbf{Y}$ given the parameters is
\begin{eqnarray}
L(\textbf{Y}|\boldsymbol\eta,\boldsymbol\Gamma_{0},\boldsymbol\Gamma_{1},\boldsymbol\Omega_{0},\boldsymbol\Omega_{1},\textbf{X})&=&|\boldsymbol\Sigma|^{-\frac{n}{2}} \exp\left\{-\frac{1}{2}tr\left[ (\textbf{Y}\boldsymbol\Gamma_{1}-\textbf{X}\boldsymbol\eta)\boldsymbol\Omega_{1}^{-1}(\textbf{Y}\boldsymbol\Gamma_{1}-\textbf{X}\boldsymbol\eta)^{T} \right]\right\}\cr
&\times& \exp\left\{-\frac{1}{2}tr\left[  \textbf{Y}\boldsymbol\Gamma_0\boldsymbol\Omega_{0}^{-1}\boldsymbol\Gamma_0^{T}\textbf{Y}^{T}\right]\right\}\cr
&=&|\boldsymbol\Sigma|^{-\frac{n}{2}} \exp\left\{-\frac{1}{2}tr\left[ \boldsymbol\Omega_{1}^{-1}(\textbf{Y}\boldsymbol\Gamma_{1} - \textbf{X}\boldsymbol\eta)^{T} (\textbf{Y}\boldsymbol\Gamma_{1}-\textbf{X}\boldsymbol\eta) \right] \right\}\cr
&&\times \exp\left\{-\frac{1}{2}tr\left[ \boldsymbol\Omega_{0}^{-1}\boldsymbol\Gamma_0^{T}\textbf{Y}^{T}\textbf{Y}^{T}\boldsymbol\Gamma_0\right]\right\}.
\end{eqnarray}
Simulating from the full conditional posterior density of $\mathbb\eta$. First, let $\mathbb\eta$ given $\boldsymbol\Gamma_{0},\boldsymbol\Gamma_{1},\boldsymbol\Omega_{0},\boldsymbol\Omega_{1}$ has $TN(\mathbb\eta_0,\Omega_1,\cdots,\Omega_m,\textbf{I}_p)$ where $\mathbb\eta_0=[\mathbbm{B},\Gamma_{11}^{T},\cdots, \Gamma_{1m}^{T},\textbf{I}_p]$.
Then, the prior distribution is
\begin{eqnarray}
\pi(\boldsymbol\eta|\boldsymbol\Gamma_{0},\boldsymbol\Gamma_{1},\boldsymbol\Omega_{0},\boldsymbol\Omega_{1})&=&|\boldsymbol\Omega_{1}|^{-\frac{p}{2}} \cr
&&\exp\left\{-\frac{1}{2}tr\left[ \boldsymbol\Omega_{1}^{-1}(\boldsymbol\eta- \mathbbm{B}_{(m+1)}\boldsymbol\Gamma_{1})^{T}( \mathbbm{B}_{(m+1)}\boldsymbol\Gamma_{1}) \right] \right\}.~~~~~~
\end{eqnarray}
Therefore, the full conditional posterior density of $\boldsymbol\eta$ given $\boldsymbol\Gamma_{0},\boldsymbol\Gamma_{1},\boldsymbol\Omega_{0},\boldsymbol\Omega_{1}$ is given by
\begin{small}
\begin{eqnarray}
&&\pi(\boldsymbol\eta|\boldsymbol\Gamma_{0},\boldsymbol\Gamma_{1},\boldsymbol\Omega_{0},\boldsymbol\Omega_{1},\textbf{Y}, \textbf{X})\cr
&\propto&\exp\left\{-\frac{1}{2}tr\left[ \boldsymbol\Omega_{1}^{-1}\left(\boldsymbol\Gamma_{1}^{T}\textbf{Y}^{T}\textbf{Y}\boldsymbol\Gamma_{1}- \boldsymbol\Gamma_{1}^{T}\textbf{Y}^{T}\textbf{X}\boldsymbol\eta- \boldsymbol\eta^{T}\textbf{X}^{T}\textbf{Y}\boldsymbol\Gamma_{1} + \boldsymbol\eta^{T}\textbf{X}^{T}\textbf{X}\boldsymbol\eta \right) \right] \right\}\cr
&\times& \exp\left\{-\frac{1}{2}tr\left[ \boldsymbol\Omega_{1}^{-1}\left(\boldsymbol\eta^{T}\boldsymbol\eta - \boldsymbol\Gamma_{1}^{T} \mathbbm{B}_{(m+1)}^{T}\boldsymbol\eta- \boldsymbol\eta^{T} \mathbbm{B}_{(m+1)}\boldsymbol\Gamma_{1} + \boldsymbol\Gamma_{1}^{T} \mathbbm{B}_{(m+1)}^{T} \mathbbm{B}_{(m+1)}\boldsymbol\Gamma_{1} \right) \right] \right\}\cr
&\propto& \exp\left\{-\frac{1}{2}tr\left[ \boldsymbol\Omega_{1}^{-1}\left(\boldsymbol\eta^{T}(\textbf{I}_{n}+\textbf{X}^{T} \textbf{X})\boldsymbol\eta - (\boldsymbol\Gamma_{1}^{T} \mathbbm{B}_{(m+1)}^{T} + \boldsymbol\Gamma_{1}^{T}\textbf{Y}^{T}\textbf{X}) \boldsymbol\eta- \boldsymbol\eta^{T}(\textbf{X}^{T} \textbf{Y}\boldsymbol\Gamma_{1}+ \mathbbm{B}_{(m+1)}\boldsymbol\Gamma_{1})  \right) \right] \right\}\cr
&\propto& \exp\left\{-\frac{1}{2}tr\left[ \boldsymbol\Omega_{1}^{-1}\left(\boldsymbol\eta^{T}(\textbf{I}_{n}+\textbf{X}^{T} \textbf{X})\boldsymbol\eta - \boldsymbol\Gamma_{1}^{T}( \mathbbm{B}_{(m+1)}^{T} + \textbf{Y}^{T}\textbf{X}) \boldsymbol\eta - \boldsymbol\eta^{T}(\textbf{X}^{T} \textbf{Y}+ \mathbbm{B}_{(m+1)}) \boldsymbol\Gamma_{1} \right) \right] \right\}.~~~
\end{eqnarray}
\end{small}
Therefore, the full conditional posterior density of $\boldsymbol\eta$ is
\begin{small}
\begin{eqnarray}
\label{posteta}
\pi(\boldsymbol\eta|\boldsymbol\Gamma_{0},\boldsymbol\Gamma_{1},\boldsymbol\Omega_{0},\boldsymbol\Omega_{1},\textbf{Y}, \textbf{X})\sim MN_{(p\times \prod_{k} u_{k})} \left((\textbf{I}+\textbf{X}^{T}\textbf{X})^{-1}(\textbf{X}^{T}\textbf{Y}+ \mathbbm{B}_{(m+1)})\boldsymbol\Gamma_{1},(\textbf{I}+ \textbf{X}^{T}\textbf{X})^{-1}, \boldsymbol\Omega_{1} \right)
\end{eqnarray}
\end{small}
It follows from (\ref{posteta}) that we can simulate from the full conditional posterior distribution of $\boldsymbol\eta$ given $\boldsymbol\Gamma_{0},\boldsymbol\Gamma_{1},\boldsymbol\Omega_{0},\boldsymbol\Omega_{1}$. Now the marginal likelihood can be calculated by integrating the $\boldsymbol\eta$. The marginal likelihood is,
\begin{footnotesize}
\begin{eqnarray}
L(\textbf{Y}|\boldsymbol\Gamma_{0},\boldsymbol\Gamma_{1},\boldsymbol\Omega_{0},\boldsymbol\Omega_{1},\textbf{X}) = |\boldsymbol\Sigma|^{-\frac{n}{2}}|\textbf{I}+ \textbf{X}^{T}\textbf{X}|^{-\frac{\prod_{k=1}^{m} r_{k}}{2}}
\exp\left\{-\frac{1}{2}tr\left[ \boldsymbol\Omega_{1}^{-1} \boldsymbol\Gamma_{1}'\textbf{G}\boldsymbol\Gamma_{1} \right] - \frac{1}{2}tr\left[  \boldsymbol\Omega_{0}^{-1}\boldsymbol\Gamma_0^{T}\textbf{Y}^{T}\textbf{Y}^{T}\boldsymbol\Gamma_0 \right]\right\}
\end{eqnarray}
\end{footnotesize}
where $\textbf{G}=\textbf{Y}^{T}\textbf{Y}+ \mathbbm{B}_{(m+1)}^{T} \mathbbm{B}_{(m+1)}+(\textbf{X}^{T}\textbf{Y}+ \mathbbm{B}_{(m+1)})^{T}(\textbf{I}+ \textbf{X}^{T}\textbf{X})^{-1}(\textbf{X}^{T}\textbf{Y}+ \mathbbm{B}_{(m+1)})$.


For simulating from the posterior density of $\boldsymbol\omega_{1k}$ given $\boldsymbol\Gamma_{0},\boldsymbol\Gamma_{1},\boldsymbol\omega_{0k}$, we let $\boldsymbol\omega_{1k}$ have the following prior distribution
\begin{eqnarray*}
\pi(\boldsymbol\omega_{1k})\propto\prod_{i_{k}=1}^{u_{k}} w_{1k,i_{k}}^{-\alpha_{k}-1}\exp\left\{-\frac{\lambda_{k}}{w_{1k,i_{k}}}\right\}.
\end{eqnarray*}
and $\textbf{M}=\boldsymbol\Gamma^{T}_1\textbf{G} \boldsymbol\Gamma_1$. Therefore, the full conditional posterior density of $\boldsymbol\omega_{1k}$ is
\begin{eqnarray}
&&\pi(\boldsymbol\omega_{1k}|\boldsymbol\Gamma_{0},\boldsymbol\Gamma_{1},\boldsymbol\Omega_{0},\boldsymbol\Omega_{1},\textbf{Y}, \textbf{X})\propto\prod_{i_{k}=1}^{u_{k}} w_{1k,i_{k}}^{-\alpha_{k}-1}\exp\left\{-\frac{\lambda_{k}}{w_{1k,i_{k}}}\right\}\cr
&\times&\prod_{i_{k}=1}^{u_{k}} w_{1k,i_{k}}^{-\frac{n}{2}\prod_{i=1,i\neq k}^{m} r_{i}}\exp\left\{-\frac{1}{2}tr(\boldsymbol\Omega_{1}^{-1}\textbf{M})\right\}\cr
&\propto&\prod_{i_{k}=1}^{u_{k}} w_{1k,i_{k}}^{-(\alpha_{k}+\frac{n\prod_{i=1,i\neq k}^{m} r_{i}}{2})-1}\exp\left\{-\frac{\lambda_{k}}{w_{1k,i_{k}}}\right\}\exp\left\{-\frac{1}{2}tr(\textbf{C}_{1k}\boldsymbol\Omega_{1}^{-1}\textbf{C}_{1k}^T\textbf{C}_{1k}\textbf{M}\textbf{C}_{1k}^T)\right\}\cr
&\propto&\prod_{i_{k}=1}^{u_{k}} w_{1k,i_{k}}^{-(\alpha_{k}+\frac{n\prod_{i=1,i\neq k}^{m} r_{i}}{2})-1}\exp\left\{-\frac{\lambda_{k}}{w_{1k,i_{k}}}\right\}\exp\left\{-\frac{1}{2}tr(\textbf{C}_{1k}\textbf{M}\textbf{C}_{1k}^T(\boldsymbol\Omega_{1}^{-k}\otimes\Omega_{1k}^{-1}))\right\}\cr
&\propto&\prod_{i_{k}=1}^{u_{k}} w_{1k,i_{k}}^{-(\alpha_{k}+\frac{n\prod_{i=1,i\neq k}^{m} r_{i}}{2})-1}\exp\left\{-\frac{\lambda_{k}}{w_{1k,i_{k}}}\right\}\exp\left\{-\frac{1}{2}tr(\tilde{\textbf{M}}_{ck}\Omega_{1k}^{-1})\right\},
\end{eqnarray}
Therefore, the posterior distribution of $\boldsymbol\omega_{1k}$ is 
\begin{eqnarray}
\label{omega1}
\pi(\boldsymbol\omega_{1k}|\boldsymbol\Gamma_{0},\boldsymbol\Gamma_{1},\boldsymbol\omega_{0k},\textbf{Y},\textbf{X})\propto \prod_{i_{k}=1}^{u_{k}} w_{1k,i_{k}}^{-(\alpha_{k}+\frac{n\prod_{i=1,i\neq k}^{m} r_{i}}{2})-1}\exp\left\{-\frac{2\lambda_{k}+\tilde{\textbf{M}}_{ck,i_ki_k}}{2w_{1k,i_{k}}}\right\}
\end{eqnarray}
where $0\leq\omega_{1k,u_k}\leq\cdots\leq\omega_{1k,1}\leq\infty$. This means, for $1 \leq j_{k}\leq u_{k}$, we can simulate from the conditional posterior density of $\boldsymbol\omega_{1k,j_{k}}$ given $\boldsymbol\Gamma_{0},\boldsymbol\Gamma_{1},\boldsymbol\omega_{0k},\boldsymbol\omega_{1k,(-j_{k})},\textbf{Y}, \textbf{X}$ by making a draw from the following truncated-Inverse-Gamma 
\[
\left(\alpha_{k}+\frac{n\prod_{i=1,i\neq k}^{m} r_{i}}{2}, \frac{2\lambda_{k}+\tilde{\textbf{M}}_{ck,j_kj_k}}{2}, \omega_{1k,j_k+1},\omega_{1k,j_k-1} \right).
\]
This distribution uses $\textbf{C}_{1k}=I_{u_m\times \cdots\times u_{k+1}}\otimes K_{u_{k-1}\times\cdots\times u_1,u_k}$.
and $\boldsymbol\Omega_{1}^{-k}=\Omega_{1m}^{-1}\otimes\cdots\otimes \Omega_{1k+1}^{-1}\otimes \Omega_{1k-1}^{-1}\otimes\cdots\otimes \Omega_{11}^{-1}=[\gamma_{1ij}^{-1}]$. Let $\textbf{M}_{ck}=\textbf{C}_{1k}\textbf{M}\textbf{C}_{1k}^{T}$,if we partition $\textbf{M}_{ck}$ to $u_k \times u_k$ matrices of $\textbf{M}_{ij}$ then $\tilde{\textbf{M}}_{ck}=\sum_{i}\sum_{j} \gamma^{-1}_{1ij}\textbf{M}_{ji}$.


For simulating from the posterior density of $\boldsymbol\omega_{0k}$ given $\boldsymbol\Gamma_{0},\boldsymbol\Gamma_{1},\boldsymbol\omega_{1k}$, we let $\boldsymbol\omega_{1k}$ have the following prior distribution
\begin{eqnarray*}
\pi(\boldsymbol\omega_{0k}|\boldsymbol\Gamma_{0},\boldsymbol\Gamma_{1},\boldsymbol\Omega_{0},\boldsymbol\Omega_{1},\textbf{Y}, \textbf{X})\propto\prod_{i_{k}=1}^{r_k-u_{k}} w_{0k,i_{k}}^{-\alpha_{0k}-1}\exp\left\{-\frac{\lambda_{0k}}{w_{0k,i_{k}}}\right\}.
\end{eqnarray*}
therefore, the full conditional posterior density is
\begin{small}
\begin{eqnarray}
&&\pi(\boldsymbol\omega_{0k}|\ldots,\textbf{Y}, \textbf{X})\propto\prod_{i_{k}=1}^{r_k-u_{k}} w_{0k,i_{k}}^{-\alpha_{0k}-1}\exp\left\{-\frac{\lambda_{0k}}{w_{0k,i_{k}}}\right\}\cr
&\times&\prod_{i_{k}=1}^{r_k-u_{k}} w_{0k,i_{k}}^{-\frac{n}{2}\prod_{i=1,i\neq k}^{m} r_{i}}\exp\left\{-\frac{1}{2}tr(\boldsymbol\Omega_{0}^{-1}\textbf{N})\right\}\cr
&\propto&\prod_{i_{k}=1}^{r_k-u_{k}} w_{0k,i_{k}}^{-\alpha_{0k}-1}\exp\left\{-\frac{\lambda_{0k}}{w_{0k,i_{k}}}\right\}
\prod_{i_{k}=1}^{r_k-u_{k}} w_{0k,i_{k}}^{-\frac{n}{2}\prod_{i=1,i\neq k}^{m} r_{i}}\exp\left\{-\frac{1}{2}tr(\textbf{C}_{0k}\boldsymbol\Omega_{0}^{-1}\textbf{C}_{0k}^T\textbf{C}_{0k}\textbf{N}\textbf{C}_{0k}^T)\right\}\cr
&\propto&\prod_{i_{k}=1}^{r_k-u_{k}} w_{0k,i_{k}}^{-\alpha_{0k}-1}\exp\left\{-\frac{\lambda_{0k}}{w_{0k,i_{k}}}\right\}
\prod_{i_{k}=1}^{r_k-u_{k}} w_{0k,i_{k}}^{-\frac{n}{2}\prod_{i=1,i\neq k}^{m} r_{i}}\exp\left\{-\frac{1}{2}tr(\textbf{C}_{0k}\boldsymbol\Omega_{0}^{-1}\textbf{C}_{0k}^T\textbf{C}_{0k}\textbf{N}\textbf{C}_{0k}^T)\right\}\cr
&\propto&\prod_{i_{k}=1}^{r_k-u_{k}} w_{0k,i_{k}}^{-\alpha_{0k}-1}\exp\left\{-\frac{\lambda_{0k}}{w_{0k,i_{k}}}\right\}
\prod_{i_{k}=1}^{r_k-u_{k}} w_{0k,i_{k}}^{-\frac{n}{2}\prod_{i=1,i\neq k}^{m} r_{i}}\exp\left\{-\frac{1}{2}tr(\textbf{C}_{0k}\textbf{N}\textbf{C}_{0k}^T(\boldsymbol\Omega_{0}^{-k}\otimes\Omega_{0k}^{-1})\right\}\cr
&\propto&\prod_{i_{k}=1}^{r_k-u_{k}} w_{0k,i_{k}}^{-\alpha_{0k}-1}\exp\left\{-\frac{\lambda_{0k}}{w_{0k,i_{k}}}\right\}
\prod_{i_{k}=1}^{r_k-u_{k}} w_{0k,i_{k}}^{-\frac{n}{2}\prod_{i=1,i\neq k}^{m} r_{i}}\exp\left\{-\frac{1}{2}tr(\tilde{\textbf{N}}_{ck}\Omega_{1k}^{-1})\right\}
\end{eqnarray}
\end{small}
where $\textbf{N} = \boldsymbol\Gamma^{T}_0\textbf{Y}^{T}(\boldsymbol\theta)\textbf{Y}\boldsymbol\Gamma_0$ and
 $\textbf{C}_{0k}=\textbf{I}_{r_m-u_m\times \cdots\times r_{k+1}-u_{k+1}}\otimes \textbf{K}_{r_{k-1}-u_{k-1}\times\cdots\times r_1-u_1,r_k-u_k}$, $\boldsymbol\Omega_0^{-k}=\boldsymbol\Omega_{0m}^{-1}\otimes\cdots\otimes \boldsymbol\Omega_{0k+1}^{-1}\otimes \boldsymbol\Omega_{0k-1}^{-1}\otimes\cdots\otimes \boldsymbol\Omega_{01}^{-1}=[\gamma^{-1}_{0ij}]$. Let $\textbf{N}_{ck}=\textbf{C}_{0k}\textbf{N}\textbf{C}_{0k}^{T}$,if we partition $\textbf{N}_{ck}$ to $r_k-u_k \times r_k-u_k$ matrices of $\textbf{N}_{ij}$ then $\tilde{\textbf{N}}_{ck}=\sum_{i}\sum_{j} \gamma^{-1}_{0ij}\textbf{N}_{ji}$.
 
Thus,
\begin{eqnarray}
\label{omega0}
\pi(\boldsymbol\omega_{0k}|\boldsymbol\Gamma_{0},\boldsymbol\Gamma_{1},\boldsymbol\omega_{1k},\textbf{Y}, \textbf{X})\propto \prod_{i_{k}=1}^{r_k-u_{k}} w_{0k,i_{k}}^{-(\alpha_{0k}+\frac{n\prod_{i=1,i\neq k}^{m} r_{i}}{2})-1}\exp\left\{-\frac{2\lambda_{0k}+\tilde{\textbf{N}}_{ck,i_ki_k}}{2w_{0k,i_{k}}}\right\}.
\end{eqnarray}
where $0\leq\omega_{0k,r_k-u_k}\leq\cdots\leq\omega_{0k,1}\leq\infty$. This means, for $1 \leq j_{k}\leq u_{k}$, we can simulate from the conditional posterior density of $\boldsymbol\omega_{0k,j_{k}}$ given $\boldsymbol\Gamma_{0},\boldsymbol\Gamma_{1},\boldsymbol\omega_{0k,(-j_{k})},\boldsymbol\omega_{1k},\textbf{Y}, \textbf{X}$ by making a draw from the following truncated-Inverse-Gamma distribution
\[
\left(\alpha_{0k}+\frac{n\prod_{i=1,i\neq k}^{m} r_{i}}{2}, \frac{2\lambda_{0k} + \tilde{\textbf{N}}_{ck,j_kj_k}}{2}, \omega_{0k,j_k+1},\omega_{0k,j_k-1} \right).
\]

To simulate from $\textbf{O}_{k} = [\boldsymbol\Gamma_{1k}~~\boldsymbol\Gamma_{0k}]$, suppose $\textbf{O}_{k,\cdot i}$ and $\textbf{O}_{k,\cdot j}$ denote the $i$th and $j$th columns of the matrix $\textbf{O}_{k}$ respectively, where $1 \leq i < j \leq r_{k}$. Let 
\begin{eqnarray*}
\pi(\textbf{O}_k|\boldsymbol\Gamma_{0}^{-k},\boldsymbol\Gamma_{1}^{-k},\boldsymbol\Omega_{0},\boldsymbol\Omega_{1}) \propto \exp\left\{-\frac{1}{2}tr(\textbf{B}_k\textbf{O}_{k}^{T}\textbf{A}_k \textbf{O}_{k})\right\}
\end{eqnarray*}
be the prior distribution for $\textbf{O}_k$ given $\boldsymbol\Gamma_{0}^{-k},\boldsymbol\Gamma_{1}^{-k},\boldsymbol\Omega_{0},\boldsymbol\Omega_{1}$. Furthermore, suppose $\textbf{O}_{k}^{-(i,j)}$ shows the matrix obtained by removing the $i$th and $j$th columns of the matrix $\textbf{O}_{k}$. Therefore, the posterior distribution can be obtain from 
\begin{small}
\begin{eqnarray}
&&\pi(\textbf{O}_k|\boldsymbol\Gamma_{0}^{-k},\boldsymbol\Gamma_{1}^{-k},\boldsymbol\Omega_{0},\boldsymbol\Omega_{1},\textbf{Y},\textbf{X})\cr
&\propto&\exp\left\{-\frac{1}{2}tr\left[ \boldsymbol\Omega_{1}^{-1} \boldsymbol\Gamma_{1}^{T}\textbf{G}\boldsymbol\Gamma_{1} \right]-\frac{1}{2}tr\left[\boldsymbol\Omega_{0}^{-1}\boldsymbol\Gamma_0^{T}\textbf{Y}^{T}\textbf{Y}\boldsymbol\Gamma_0\right]\right\}\exp\left\{-\frac{1}{2}tr(\textbf{B}_k\textbf{O}_{k}^{T}\textbf{A}_k \textbf{O}_{k})\right\}~~~~~~
\end{eqnarray}
\end{small}
Using the following algebra, one can commute the Kronecker product and rewrite the $tr\left[ \boldsymbol\Omega_{1}^{-1} \boldsymbol\Gamma_{1}^{T}\textbf{G}\boldsymbol\Gamma_{1} \right]$ as follows
\begin{eqnarray}
&&tr\left[ \boldsymbol\Omega_{1}^{-1} \boldsymbol\Gamma_{1}^{T}\textbf{G}\boldsymbol\Gamma_{1} \right]=tr\left[\textbf{G}\boldsymbol\Gamma_{1} \boldsymbol\Omega_{1}^{-1} \boldsymbol\Gamma_{1}^{T} \right]=tr\left[\textbf{C}_k\textbf{G}\textbf{C}_k^{T}\textbf{C}_k\boldsymbol\Gamma_{1} \boldsymbol\Omega_{1}^{-1} \boldsymbol\Gamma_{1}^{T}\textbf{C}_k^{T}\right]\cr
&=&tr\left[\textbf{C}_k\textbf{G}\textbf{C}_k^{T}\textbf{C}_k\boldsymbol{V}_1^\dagger \textbf{C}_k^{T}\right]=tr\left[\textbf{C}_k\textbf{G}\textbf{C}_k^{T}(\textbf{V}_1^{k\dagger} \otimes \textbf{V}_{1k}^\dagger)\right],
\end{eqnarray}
where 
$\boldsymbol{V}_1^{k\dagger}=\textbf{V}_{1m}^{k\dagger}\otimes\cdots\otimes \textbf{V}_{1k+1}^{k\dagger}\otimes \textbf{V}_{1k-1}^{k\dagger}\otimes\cdots\otimes \textbf{V}_{11}^{k\dagger}=[v_{1ij}^{k\dagger}]$.
Let $\textbf{G}_{ck}=\textbf{C}_k\textbf{G}\textbf{C}_k^{T}$,if we partition $\textbf{G}_{ck}$ to $r_k \times r_k$ matrices of $\textbf{G}_{ij}$ then $\tilde{\textbf{G}}_{ck}=\sum_{i}\sum_{j} v^{k\dagger}_{1ij}\textbf{G}_{ji}$ and
\begin{eqnarray*}
&&tr\left[ \boldsymbol\Omega_{1}^{-1} \boldsymbol\Gamma_{1}^{T}\textbf{G}\boldsymbol\Gamma_{1} \right]=tr\left[\textbf{G}_{ck}(\textbf{V}_1^{k\dagger} \otimes \textbf{V}_{1k}^\dagger)\right]=tr\left[\tilde{\textbf{G}}_{ck}\textbf{V}_{1k}^\dagger\right]\cr
&&=tr\left[\tilde{\textbf{G}}_{ck}\boldsymbol\Gamma_{1k}\boldsymbol\Omega_{1k}^{-1} \boldsymbol\Gamma_{1k}^{T}\right]=tr\left[\boldsymbol\Omega_{1k}^{-1} \boldsymbol\Gamma_{1k}^{T} \tilde{\textbf{G}}_{ck}\boldsymbol\Gamma_{1k}\right].
\end{eqnarray*}
Define $\textbf{H}=\textbf{Y}^{T}\textbf{Y}$, then we can show:
\begin{eqnarray}
&&tr\left[ \boldsymbol\Omega_{0}^{-1} \boldsymbol\Gamma_{0}^{T}\textbf{H}\boldsymbol\Gamma_{0} \right]=tr\left[\textbf{H}_{ck}(\boldsymbol{V}_0^{k\dagger} \otimes \textbf{V}_{0k}^\dagger)\right]=tr\left[\tilde{\textbf{H}}_{ck}\textbf{V}_{0k}^\dagger\right]\cr
&&=tr\left[\tilde{\textbf{\textbf{H}}}_{ck}\boldsymbol\Gamma_{0k}\boldsymbol\Omega_{0k}^{-1} \boldsymbol\Gamma_{0k}^{T}\right]=tr\left[\boldsymbol\Omega_{0k}^{-1} \boldsymbol\Gamma_{0k}^{T} \tilde{\textbf{H}}_{ck}\boldsymbol\Gamma_{0k}\right],
\end{eqnarray}
where 
$\textbf{V}_0^{k\dagger}=\textbf{V}_{0m}^{k\dagger}\otimes\cdots\otimes \textbf{V}_{0k+1}^{k\dagger}\otimes \textbf{V}_{0k-1}^{k\dagger}\otimes\cdots\otimes \textbf{V}_{01}^{k\dagger}=[v^{k\dagger}_{0ij}]$. Let $\textbf{H}_{ck}=\textbf{C}_{k}\textbf{H}\textbf{C}_{k}^{T}$,if we partition $\textbf{H}_{ck}$ to $r_k \times r_k$ matrices of $\textbf{H}_{ij}$ then $\tilde{\textbf{H}}_{ck}=\sum_{i}\sum_{j} v^{k\dagger}_{0ij}\textbf{H}_{ji}$.
\begin{small}
\begin{eqnarray}
\label{posto}
&&\pi(\textbf{O}_k|\boldsymbol\Gamma_{0}^{-k},\boldsymbol\Gamma_{1}^{-k},\boldsymbol\Omega_{0},\boldsymbol\Omega_{1},\textbf{Y},\textbf{X})\cr
&\propto&\exp\left\{-\frac{1}{2}tr\left[\boldsymbol\Omega_{1k}^{-1} \boldsymbol\Gamma_{1k}^{T} \tilde{\textbf{G}}_{ck}\boldsymbol\Gamma_{1k}\right]-\frac{1}{2}tr\left[\boldsymbol\Omega_{0k}^{-1} \boldsymbol\Gamma_{0k}^{T} \tilde{\textbf{H}}_{ck}\boldsymbol\Gamma_{0k}\right]\right\}\exp\left\{-\frac{1}{2}tr(\textbf{B}_k\textbf{O}_{k}^{T}\textbf{A}_k \textbf{O}_{k})\right\}\cr
&=&\exp\left\{-\frac{1}{2}\sum_{i_k=1}^{u_k}\textbf{O}_{\cdot i_k}^{T}(\omega^{-1}_{1,i_k}\tilde{\textbf{G}}_{ck}+b_{i_k,i_k}\textbf{A}_k)\textbf{O}_{\cdot i_k}-\frac{1}{2}\sum_{j_k=u_k+1}^{r_k}\textbf{O}_{\cdot j_k}^{T}(\omega^{-1}_{0,j_k-u_k}\tilde{\textbf{H}}_{ck}+b_{j_k,j_k}\textbf{A}_k)\textbf{O}_{\cdot j_k}\right\},~~~~~
\end{eqnarray}
\end{small}
where $\textbf{O}_k^{T}\textbf{O}_k=\textbf{I}$. We used the algorithm proposed by \cite{khare2017bayesian} to draw samples from the posterior distribution. 

\end{document}